   \documentclass[12pt]{article}
   \usepackage{latexsym}
\else\oddsidemargin25mm\evensidemargin25mm\marginparwidth25mm\fi%
\begin{document}
%%%%%%%%%%%%%%%%%%%%%%%%%%%%%%%%%%%%%%%%%%
%%%%%%%%%%%%%%%%%%%%%%%%%%%%%%%%%%%%%%%%%%
% %%%%%%%%%%%%%%%%%%%%%%%%%%%%%%%%%%%%%%%%%
%%%%%% ANNAMACRO %%%%%%%%%%%%%%%%%%%%%%%%%%
%%%%%%%%%%%%%%%%%%%%%%%%%%%%%%%%%%%%%%%%%%%
%%%%%%%%%%%%%%%%%%%%%%%%%%%%%%%%%%%%%%%%%%%
\newcommand{\ft}[2]{{\textstyle\frac{#1}{#2}}}
\newcommand{\QED}{{\hspace*{\fill}\rule{2mm}{2mm}\linebreak}}
\def\ii{{\rm i}}
\def\dop{{\rm d}\hskip -1pt}
\def\bfone{\relax{\rm 1\kern-.35em 1}}
\def\bfzero{\relax{\rm I\kern-.18em 0}}
\def\inbar{\vrule height1.5ex width.4pt depth0pt}
\def\IC{\relax\,\hbox{$\inbar\kern-.3em{\rm C}$}}
\def\ID{\relax{\rm I\kern-.18em D}}
\def\IF{\relax{\rm I\kern-.18em F}}
\def\IK{\relax{\rm I\kern-.18em K}}
\def\IH{\relax{\rm I\kern-.18em H}}
\def\II{\relax{\rm I\kern-.17em I}}
\def\IN{\relax{\rm I\kern-.18em N}}
\def\IP{\relax{\rm I\kern-.18em P}}
\def\IQ{\relax\,\hbox{$\inbar\kern-.3em{\rm Q}$}}
\def\IR{\relax{\rm I\kern-.18em R}}
\def\IG{\relax\,\hbox{$\inbar\kern-.3em{\rm G}$}}
\font\cmss=cmss10 \font\cmsss=cmss10 at 7pt
\def\ZZ{\relax\ifmmode\mathchoice
{\hbox{\cmss Z\kern-.4em Z}}{\hbox{\cmss Z\kern-.4em Z}}
{\lower.9pt\hbox{\cmsss Z\kern-.4em Z}} {\lower1.2pt\hbox{\cmsss
Z\kern .4em Z}}\else{\cmss Z\kern-.4em Z}\fi}
\def\a{\alpha} \def\b{\beta} \def\d{\delta}
\def\e{\epsilon} \def\c{\gamma}
\def\G{\Gamma} \def\l{\lambda} \def\g{\gamma}
\def\L{\Lambda} \def\s{\sigma} \def\l{\lambda} \def\p{\psi} \def\pb{\overline{\psi}}
\def\lb{\overline{\lambda}}
\def\cA{{\cal A}} \def\cB{{\cal B}}
\def\cC{{\cal C}} \def\cD{{\cal D}}
    \def\cF{{\cal F}} \def\cG{{\cal G}}
\def\cH{{\cal H}} \def\cI{{\cal I}}
\def\cJ{{\cal J}} \def\cK{{\cal K}}
\def\cL{{\cal L}} \def\cM{{\cal M}}
\def\cN{{\cal N}} \def\cO{{\cal O}}
\def\cP{{\cal P}} \def\cQ{{\cal Q}}
\def\cR{{\cal R}} \def\cV{{\cal V}}\def\cW{{\cal W}}
%
%
%%%%%%%%%%%%%%%%%%%%%%%%%%%%%%%%%%%%%%%%%%%%%%%%%%%%%%%%%%%%%%
%%%%% misc macros %%%%%
%%%%%%%%%%%%%%%%%%%%%%%%%%%%%%%%%%%%%%%%%%%%%%%%%%%%%%%%%%%%%%
%
\def\crr{\crcr\noalign{\vskip {8.3333pt}}}
\def\tilde{\widetilde}
\def\bar{\overline}
\def\us#1{\underline{#1}}
\let\shat=\hat
\def\hat{\widehat}
\def\hyp{\vrule height 2.3pt width 2.5pt depth -1.5pt}
\def\square{\mbox{.08}{.08}}
\def\eq#1{(\ref{#1})}
\def\Coeff#1#2{{#1\over #2}}
\def\Coe#1.#2.{{#1\over #2}}
\def\coeff#1#2{\relax{\textstyle {#1 \over #2}}\displaystyle}
\def\coe#1.#2.{\relax{\textstyle {#1 \over #2}}\displaystyle}
\def\half{{1 \over 2}}
\def\shalf{\relax{\textstyle {1 \over 2}}\displaystyle}
\def\dag#1{#1\!\!\!/\,\,\,}
\def\to{\rightarrow}
\def\notin{\hbox{{$\in$}\kern-.51em\hbox{/}}}
\def\shdot{\!\cdot\!}
\def\ket#1{\,\big|\,#1\,\big>\,}
\def\bra#1{\,\big<\,#1\,\big|\,}
\def\equaltop#1{\mathrel{\mathop=^{#1}}}
\def\Trbel#1{\mathop{{\rm Tr}}_{#1}}
\def\inserteq#1{\noalign{\vskip-.2truecm\hbox{#1\hfil}
\vskip-.2cm}}
\def\attac#1{\Bigl\vert
{\phantom{X}\atop{{\rm\scriptstyle #1}}\phantom{X}}}
\def\exx#1{e^{{\displaystyle #1}}}
\def\del{\partial}
\def\delbar{\bar\partial}
\def\nex#1{$N\!=\!#1$}
\def\dex#1{$d\!=\!#1$}
\def\cex#1{$c\!=\!#1$}
\def\eg{{\it e.g.}} \def\ie{{\it i.e.}}
%\catcode`\@=12
%%%%%%%%%%%%%%%%%%%%%%%%%%%%%%%%%%%%%%%%%%%%%%%%%%%%%%%%%%%%%%
%%%%%%%%%%%%%%%%%%%%%%%%%%%%%%%%%%%%%%%%%%%%%%%%%%%%%%%%%%%%
%\draft
%%%%%%%%%%%% macros and references %%%%%%%%%%%%%%%%%%%%%%%%%
%
\def\cS{{\cal K}}
\def\IE{\relax{{\rm I\kern-.18em E}}}
\def\cE{{\cal E}}
\def\rt{{\cR^{(3)}}}
\def\IGam{\relax{{\rm I}\kern-.18em \Gamma}}
\def\IGa{\IA}
\def\LG{Lan\-dau-Ginz\-burg\ }
\def\cV{{\cal V}}
\def\Rt{{\cal R}^{(3)}}
\def\wabc{W_{abc}}
\def\WABC{W_{\a\b\c}}
\def\W{{\cal W}}
\def\tft#1{\langle\langle\,#1\,\rangle\rangle}
\def\IA{\relax{\hbox{{\rm A}\kern-.82em {\rm A}}}}
\let\picfuc=\fp
\def\hata{{\shat\a}}
\def\hatb{{\shat\b}}
\def\hatA{{\shat A}}
\def\hatB{{\shat B}}
\def\bv{{\bf V}}
\def\spg{special geometry}
\def\sc{SCFT}
\def\leel{low energy effective Lagrangian}
\def\pf{Picard--Fuchs}
\def\pfS{Picard--Fuchs system}
\def\el{effective Lagrangian}
\def\Fb{\overline{F}}
\def\nablab{\overline{\nabla}}
\def\Ub{\overline{U}}
\def\Db{\overline{D}}
\def\zb{\overline{z}}
\def\eb{\overline{e}}
\def\fb{\overline{f}}
\def\tb{\overline{t}}
\def\Xb{\overline{X}}
\def\Vb{\overline{V}}
\def\Cb{\overline{C}}
\def\Sb{\overline{S}}
\def\delb{\overline{\del}}
\def\Gammab{\overline{\Gamma}}
\def\Ab{\overline{A}}
\def\Anh{A^{\rm nh}}
\def\alphab{\bar{\alpha}}
\def\cy{Calabi--Yau}
\def\cabg{C_{\alpha\beta\gamma}}
\def\B{\Sigma}
\def\Bh{\hat \Sigma}
\def\Kh{\hat{K}}
\def\Knh{{\cal K}}
\def\A{\Lambda}
\def\Ah{\hat \Lambda}
\def\R{\hat{R}}
\def\V{{V}}
\def\T{T}
\def\Gammah{\hat{\Gamma}}
\def\twot{$(2,2)$}
\def\K{K\"ahler}
\def\rat{({\theta_2 \over \theta_1})}
\def\lv{{\bf \omega}}
\def\w{w}
\def\CP{C\!P}
\def\o#1#2{{{#1}\over{#2}}}
%%%%%%%%%%%%%%%%%%%%%%%%%%%%%%%%%%%%%%%%%%%%%%%%%%%%%%%
% FINE DELLA ANNAMACRO.TEX %%%%%%%%%%%%%%%%%%%%%%%%%%%%
%%%%%%%%%%%%%%%%%%%%%%%%%%%%%%%%%%%%%%%%%%%%%%%%%%%%%%%
%%%%%%%%%%%%%%%%%%%%%%%%%%%%%%%%%%%%%%%%%%%%%%%%%%%%%%%
% Paolo Macro %%%%%%%%%%%%%%%%%%%%%%%%%%%%%%%%%%%%%%%%%
%%%%%%%%%%%%%%%%%%%%%%%%%%%%%%%%%%%%%%%%%%%%%%%%%%%%%%%
\newcommand{\be}{\begin{equation}}
\newcommand{\ee}{\end{equation}}
\newcommand{\ba}{\begin{eqnarray}}
\newcommand{\ea}{\end{eqnarray}}
\newcommand{\brr}{\begin{array}}
\newcommand{\err}{\end{array}}
\newcommand{\nn}{\nonumber}
%%%%%%%%%%%%%%%%%%%%%%%%%%%%%%%%%%%%%%%%%%%%%%
\def\twomat#1#2#3#4{\left(\begin{array}{cc}
 {#1}&{#2}\\ {#3}&{#4}\\
\end{array}
\right)}
\def\twovec#1#2{\left(\begin{array}{c}
{#1}\\ {#2}\\
\end{array}
\right)}
%%%%%%%%%%%%%%%%%%%%%%%%%%%%%%%%%%%%%%%%%%%%%%%%%%%%
%%%%%%%%%%%%%%%%%%%%%%%%%%%%%%%%%%%%%%%%%%%%%%%%%%%%%%%%%%%%%
%% This File is the titlepage %%%%%%%%%%%%%%%%%%%%%%%%%%%%%%%%
%%%%%%%%%%%%%%%%%%%%%%%%%%%%%%%%%%%%%%%%%%%%%%%%%%%%%%%%%%%%%
\begin{titlepage}
\hskip 5.5cm
%\vbox{
%\hbox{CERN-TH/2000-167}
%}
\hskip 5.5cm \hbox{April, 2001} \vfill \vskip 2cm
\begin{center}
{\LARGE {Matter Coupled $F(4)$ Gauged Supergravity Lagrangian}}\\
%\vfill
\vskip 1.5cm
  {\bf Laura Andrianopoli, Riccardo D'Auria and
Silvia Vaul\`a} \\
%\vfill
\vskip 0.5cm {\small Dipartimento di Fisica, Politecnico di
Torino,\\
 Corso Duca degli Abruzzi 24, I-10129 Torino\\
and Istituto Nazionale di Fisica Nucleare (INFN) - Sezione di
Torino, Italy}\\
\vspace{6pt}
\end{center}
\vskip 3cm
\begin {abstract}
We construct the so far unknown Lagrangian of $D=6$, $N=2$ $F(4)$
Supergravity coupled to an arbitrary number of vector multiplets
whose scalars span the coset manifold $\frac{SO(4,n)}{SO(4)\times
SO(n)}$.
 This is done first in the ungauged case and then extended to the
 compact
gauging of $SU(2)\times {\mathcal G}$, where $SU(2)$ is the
$R$-symmetry diagonal subgroup of $SU(2)_L \times SU(2)_R \simeq
SO(4)$ and ${\mathcal G}$ is a compact subgroup of $SO(n)$, $n$
being the number of vector multiplets, and such that
dim(${\mathcal G}$) $= n$.
 The knowledge of the Lagrangian allows in principle to
refine the $AdS_6/CFT_5$ correspondence already discussed, as far
as supersymmetric multiplets are concerned, in a previous related
paper. With respect to the latter we also give a more exaustive
treatment of the construction of the theory at the level of
superspace Bianchi identities and in particular of the scalar
potential.
\end{abstract}
\vfill
%{\small
%\scriptsize
%abstract}

\end{titlepage}
%%%%%%%%%%%%%%%%%%%%%%%%%%%%%%%%%%%%%%%%%%%%%%%%%%%%%%%%%%%%%%%%%%%%%%
%%%%%%%%%%%%               INTRODUZIONE                %%%%%%%%%%%%%%%%
%%%%%%%%%%%%%%%%%%%%%%%%%%%%%%%%%%%%%%%%%%%%%%%%%%%%%%%%%%%%%%%%%%%%%%%
\section{Introduction}

In the classical papers of references \cite{nrs}, \cite{Kac} it
was shown that, in the classification of Lie superalgebras, there
appear a few exceptional superalgebras in analogous way to what
happens in the Cartan classification of Lie algebras. Among these
we find a $F(4)$ superalgebra whose name is due to the fact that
its construction can be realized starting from the $F(4)$ Lie
algebra with a standard procedure. This superalgebra is in fact
the minimal extension in $D=5$ of the conformal group $SO(2,5)$ or
equivalently of the Anti de Sitter ($AdS$) group in six
dimensions\footnote{There exist a non minimal supersymmetric
extension of the same group giving the orthosymplectic group
$Osp(8^*|2)$ \cite{dflv}}. As we will see in the following, the
$F(4)$ superalgebra contains as maximal bosonic subalgebra
$so(2,5)\otimes su(2)$, so it is the natural candidate for the
construction of a supergravity theory in six dimensions with 16
supersymmetries ($N=2$, $D=6$ supergravity). This theory was
indeed constructed, without coupling to matter fields, in
reference \cite{rom} and, as it was to be expected, it exhibits an
$AdS_6$ supersymmetric background when a particular relation
between the $AdS_6$ radius $R$ and the coupling constant $g$ of
the gauged $R$-symmetry $SU(2)$ occurs. This result was in fact
retrieved after the Lagrangian and the transformation rules were
constructed and the extremum of the scalar potential, depending
only on the dilaton field, was found. It was then realized $a\,\
posteriori$ that the $SU(2)$ gauged $N=2$, $D=6$ supergravity
should be the looked for $F(4)$ supergravity.\\
 In reference \cite{noi} we
extended the construction to the coupling with vector multiplets
(the only kind of supermatter in $D=6,\,\ N=2$ supergravity) in
view of the analysis of the $AdS/CFT$ correspondence between
$F(4)$ matter coupled supergravity and the 5-dimensional
superconformal field theory at the fixed point of the
renormalization group, where the latter theory turns out to be a
theory of interacting 5-dimensional hypermultiplets \cite{smi},
\cite{fkpz}, \cite{oz}. Our starting point, however, was
different from Romans approach in that we constructed the theory
directly from the $F(4)$ superalgebra or better from its dual
formulation in terms of Maurer-Cartan equations (M.C.E.) suitably
extended to a Free Differential Algebra (F.D.A.). In particular
in this approach the relation between the $SU(2)$ gauge coupling
constant and the $AdS$ radius appears as a natural consequence of
the $F(4)$ superalgebra structure constants. Furthermore, when
the M.C.E. are extended to a F.D.A in order to include all the
fields of the supergravity supermultiplet, the dynamical Higgs
mechanism found by Romans, through which the antisymmetric tensor
field becomes massive by eating the $SU(2)$ gauge field singlet,
turns out to be an obvious consequence of
the structure of the F.D.A.\\
Since in reference \cite{noi} we were mainly interested in the
construction of the $AdS/CFT$ correspondence, the focus of that
paper was concentrated on the construction of the supersymmetry
transformation rules by solving Bianchi identities in superspace
and in the subsequent construction of the scalar potential for the
matter coupled theory.

In this paper we want to complete the construction of the theory
by determining the matter coupled Lagrangian (up to 4 fermions
terms) and to give some more details on the geometrical
construction of the matter coupled gauge theory and on many
results which were only outlined in reference \cite{noi}.

The plan of the paper is as follows:\\
In section 2 we give the algebraic backgrounds of the $F(4)$
superalgebra together with M.C.E.'s and the extended F.D.A.,
explaining how in this framework the previously described results
of Higgs mechanism and the $g$ versus $R_{AdS}$ relation arise. \\
In section 3 we construct the matter coupled theory in absence of
gauging. We do that starting from the definition of the superspace
``curvatures" and solving the related Bianchi identities for the
matter coupled ungauged theory. The superspace Bianchi identities
solutions are then translated as ordinary transformation laws of
the physical fields on space time. The ungauged Lagrangian is
then constructed by using the geometrical approach (rheonomic) in
supespace and then restricting its form to ordinary space--time.
\\
In section 4 we perform the gauging procedure and explain how the
transformation laws of the physical fields are modified both by
the presence of the gauging and by turning on a mass parameter $m$
for the antisymmetric tensor field. This is sufficient to
conclude for the existence of an Anti de-Sitter supersymmetric
background when $g=3m \equiv 3 \left(2R_{AdS} \right)^{-1}$. \\
In section 5 the space-time Lagrangian of the gauged theory is
given and
 its properties discussed. In particular, we compute the scalar potential
 and verify the $AdS_6/CFT_5$ correspondence as far as the masses of the
 vector and gravitational multiplets are concerned. \\
 Section 6 contains our conclusions.\\
 The three Appendices A, B and C contain technical material which
 is however quite essential for achieving our results.
 In particular, in Appendix A and B we explain in some detail the
 superspace approach to the solution of Bianchi identities and to the construction of
 the Lagrangian using
 the so called ``rheonomic formalism''. Note that while the
 rheonomic approach is fully equivalent to the ordinary superspace
 approach at the level of Bianchi identities,
 the construction of the Lagrangian in the rheonomic approach has
 no parallel in the ordinary superspace formalism.
\\
Appendix C contains the relevant Fierz identities for the
construction of the theory.
%%%%%%%%%%%%%%%%%%%%%%%%%%%%%%%%%%%%%%%%%%%%%%%%%%%%%%%%%%
%%%%%%%%%%%%%%%%%%%%  physical background %%%%%%%%%%%%%%%
%%%%%%%%%%%%%%%%%%%%%%%%%%%%%%%%%%%%%%%%%%%%%%%%%%%%%%%%%%
\section{Geometrical and Physical backgrounds}
\label{geomphysback}
 \setcounter{equation}{0}
 In this section we introduce the
geometrical and physical settings
for the construction of the theory.\\
Let us recall the content of the $D=6$, $N=(1,1)$ supergravity
multiplet in a Poincar\'e background:
\begin{equation}
(V^{a}_{\mu}, A^{\alpha}_{\mu}, B_{\mu\nu},\, \psi^{A}_{\mu},\,
\psi^{\dot{A}}_{\mu}, \chi^{A}, \chi^{\dot{A}}, e^{\sigma})
\end{equation}
\noindent where $V^a_{\mu}$ is the six dimensional vielbein,
$\psi^{A}_{\mu},\,\ \psi^{\dot{A}}_{\mu}$ are left-handed and
right-handed four- component gravitino fields respectively, $A$
and $\dot{A}$ transforming under the two factors of the
$R$-symmetry group $SO(4) \simeq SU(2)_L\otimes SU(2)_R$;
$B_{\mu\nu}$ is a 2-form, $A^{\alpha}_{\mu}$ ($\alpha=0,1,2,3$),
are  vector fields, $\chi^{A}, \chi^{\dot{A}}$ are left-handed and
right-handed spin $\frac{1}{2}$ four components dilatinos, and
$e^{\sigma}$ denotes the dilaton.\\ Our notations are as follows:
$a,b,\dots=0,1,2,3,4,5$ are Lorentz flat indices in $D=6$
$\mu,\nu,\dots=0,1,2,3,4,5$ are the corresponding world indices,
  $A,\dot{A}=1,2$. Moreover our metric is
$(+,-,-,-,-,-)$.\\ We recall that the description of the spinors
of the multiplet in terms of left-handed and right-handed
projection holds only in a Poincar\'e background, while in an
$AdS$ background the chiral projection cannot be defined and we
are bounded to use 8-dimensional pseudo-Majorana spinors. Indeed
for $SO(1,5)$ (which corresponds to $D=6$, $\rho=4$, $\rho$ being
the signature of $SO(5,1)$ mod 8, in the notations of reference
\cite{dflv}), the spinors are 4-dimensional Weyl-quaternionic,
while for $SO(2,5)$ (corresponding to $D=7$, $\rho=-3$), the
spinors are 8-dimensional real-quaternionic.\footnote{Here, by
"quaternionic" we mean that they satisfy a pseudo-Majorana
condition.} In the former case the $R$-symmetry group is
$SU(2)_L\otimes SU(2)_R$, while in the latter case it reduces to
the $SU(2)$ diagonal subgroup of $SU(2)_L\otimes SU(2)_R$. For
our purposes, it is convenient to use from the very beginning
8-dimensional pseudo-Majorana spinors even in a Poincar\'e
framework, since we are going to
discuss in a unique setting both Poincar\'e and $AdS$ vacua.\\
The pseudo-Majorana condition on the gravitino 1-forms is as
follows:
\begin{equation}
(\psi_A)^{\dagger}\gamma^0=\overline{(\psi_A)}=\epsilon^{AB}\psi_{B}^{\
\ t}
\end{equation}
\noindent where we have chosen the charge conjugation matrix in
six dimensions as the identity matrix (an analogous definition
holds for the dilatino fields). We use eight dimensional
antisymmetric gamma matrices, with
$\gamma^7=i\gamma^0\gamma^1\gamma^2\gamma^3\gamma^4\gamma^5$,
which implies $\gamma_7^T=-\gamma_7$ and $(\gamma_7)^2=-1$. The
indices $A,B,\dots=1,2,\cdots $ of the spinor fields $\psi_A,\,
\chi_A$ transform in the fundamental of the diagonal subgroup
$SU(2)$ of $SU(2)_L\otimes SU(2)_R$. For a generic $SU(2)$ tensor
$T$, raising and lowering of indices are defined by
\begin{eqnarray}
&&T^{\dots A\dots}=\epsilon^{AB}\ \ T^{\dots\ \ \dots}_{\ \ B}\\
&&T_{\dots A\dots}=T_{\dots\ \ \dots}^{\ \ B}\ \ \epsilon_{BA}
\end{eqnarray}

Taking into account that the $F(4)$ supergroup has as bosonic
subgroup $SO(2,5)\otimes SU(2)$ we consider the 1-forms associated
to $AdS_6$ algebra, namely $\omega^{ab}$, $V^a$ dual to the
$SO(2,5)$ generators $M_{ab}$ and $P_a$ respectively, which
satisfy:
\begin{eqnarray}&&[M_{ab},M_{cd}]=\frac{1}{2}(\eta_{bc}M_{ad}+\eta_{ad}M_{bc}-
\eta_{bd}M_{ac}-\eta_{ac}M_{bd}) \nonumber\\
&&[P_a,P_b]=8m^2M_{ab}\nonumber\\
&&[M_{ab},P_c]=\frac{1}{2}(\eta_{ac}P_b-\eta_{bc}P_a)
\end{eqnarray}
\noindent and the 1-form $A^r$, $r=1,2,3$ dual to the $SU(2)$
generators $T_r$, satisfying: \be [T^s;T^t]=ig\epsilon^{str}T_r
\ee \noindent where $g$ is the coupling constant of $SU(2)$, and
$m$
is related to the $AdS_6$ radius of $SO(2,5)/SO(1,5)$ by $m=(2R_{AdS})^{-1}$.\\
In order to construct the full $F(4)$ superalgebra we now
introduce the pseudo-Majorana spinor charges $Q_{A\alpha}$,
($\alpha$ being an 8-dimensional spinor index) and try to enlarge
the $SO(2,5)\otimes SU(2)$ algebra to the full $F(4)$
superalgebra. The simplest procedure is to enlarge the M.C.E. of
$SO(2,5)\otimes SU(2)$ given by:
\begin{eqnarray}
&&\mathcal{D}V^{a}\equiv dV^a-\omega^{ab}V_b=0\nonumber\\
&&\mathcal{R}^{ab}+4m^{2}\ \
V^{a}V^{b}=0\nonumber\\
&&dA^{r}+\frac{1}{2} g \epsilon^{rst}A_{s}A_{t}=0
\label{mauc}
\end{eqnarray}
\noindent where $\mathcal{R}^{ab}\equiv
d\omega^{ab}-\omega^{ac}\land\,\ \omega_c^{\,\ b}$, in terms of
the  the spinor 1-forms $\p^{A\alpha}$ dual to the odd generators
$Q_{A\a}$ . It turns out that the minimal extension of
(\ref{mauc}) is given by:
\begin{eqnarray}
&&\mathcal{D}V^{a}-\frac{i}{2}\ \
\overline{\psi}_{A}\gamma_{a}\psi^A=0\nonumber\\
&&\mathcal{R}^{ab}+4m^{2}\ \
V^{a}V^{b}+m\overline{\psi}_{A}\gamma_{ab}\psi^A=0\nonumber\\
&&dA^{r}+\frac{1}{2}\,\ g\,\ \epsilon^{rst}A_{s}A_{t}-i\,\
\overline{\psi}_{A}\psi_{B}\,\
\sigma^{rAB}=0\nonumber\\
&&D\psi_{A}-im\gamma_{a}\psi_{A}V^{a}=0
\label{maucc}
\end{eqnarray} \noindent where $D$ is the
$SO(1,5)\otimes SU(2)$ covariant derivative, which on spinors acts
as follows:
\begin{equation}D\psi_A\equiv\mathcal{D}\p_A-\frac{i}{2}\sigma_{AB}^r
A_r\psi^B=
d\psi_A-\frac{1}{4}\gamma_{ab}\omega^{ab}\psi_A-\frac{i}{2}\sigma_{AB}^r
A_r\psi^B .
\end{equation}
Moreover $\sigma^{rAB}=\epsilon^{BC}\sigma^{rA}_{\ \ C}$, where
$\sigma^{rA}_{\ \ B}$\ \ ($r=1,2,3$) denote the usual Pauli
matrices, are symmetric in $A,\,\ B$.
\\
Note that equations (\ref{maucc}) are closed under
d-differentiation if and only if $g=3m$. To recover this result
one has to use the following Fierz identity involving 3-$\psi_A$'s
1-forms:
\begin{equation}
\label{fond}\frac{1}{4}\gamma_{ab}\psi_A\overline{\psi}_B\gamma^{ab}\psi_C\epsilon^{AC}-\frac{1}{2}
\gamma_{a}\psi_A\overline{\psi}_B\gamma^{a}\psi_C\epsilon^{AC}+3\psi_C\overline{\psi}_B\psi_A\epsilon^{BC}=0
\end{equation}
\noindent This identity is just one example of the many Fierz
identities necessary for the subsequent construction of the
theory. We give an account of their derivation in appendix C. At
this point the Lie Algebra (anti) commutators of the $F(4)$
supergroup are easily retrieved using the well known identity \be
d\omega (X,Y)=\frac{1}{2}\{X(\omega(Y))-Y(\omega(X))-\omega[X,Y]\}
\ee
 \noindent and the duality relations given by
\be  \omega^{ab}(M_{cd})=\d^{ab}_{cd}\ \ V^a(P_b)=\d^a_b\ \
\p^{A\a}(Q_{B\b})=\d^a_b\d^{\a}_{\b} \ee
\noindent all the other duality relations being zero.\\
The resulting $F(4)$ Lie superalgebra is:
\begin{eqnarray}&&[M_{ab},M_{cd}]=\frac{1}{2}(\eta_{bc}M_{ad}+\eta_{ad}M_{bc}-\eta_{bd}M_{ac}-\eta_{ac}M_{bd})\nonumber\\
&&[P_a,P_b]=8m^2M_{ab} \nonumber\\
&&[M_{ab},P_c]=\frac{1}{2}(\eta_{ac}P_b-\eta_{bc}P_a)
\nonumber\\ &&[T^s;T^t]=ig\epsilon^{str}T_r\nonumber\\
&&[M_{ab},\overline{Q}_{A\beta}]=-\frac{1}{4}\overline{Q}_{A\alpha}(\gamma_{ab})_{\alpha\beta}
\nonumber\\
&&[P_a,\overline{Q}_{A\beta}]=im\overline{Q}_{A\alpha}(\gamma_a)_{\alpha\beta}
\nonumber\\
&&[T_{(AB)},\overline{Q}_{C\alpha}]=\frac{i}{2}g(\overline{Q}_{A\alpha}\delta_{BC}+\overline{Q}_{B\alpha}\delta_{AC})
\nonumber\\
&&\{\overline{Q}_{A\alpha},Q_{B\beta}\}=-i\epsilon_{AB}(\gamma^a)_{\alpha\beta}P_a+4i(1)_{\alpha\beta}T_{(AB)}
+m\epsilon_{AB}(\g^{ab})_{\a\b}M_{ab}
\end{eqnarray}
\noindent where we have defined $T_{(AB)}\equiv T_r\s^r_{AB}$ and
where $g=3m$ in this Lie--superalgebra setting is now an outcome
of (super) Jacobi identities.
\par
Let us now come back to the M.C.E.'s (\ref{maucc}): note that they
keep exactly the same form if we pass from the $F(4)$ supergroup
to the quotient $F(4)/SO(1,5)\otimes SU(2)$, which is the relevant
superspace having as bosonic subcoset $AdS_6$. Once the pull-back
is done, the 1-forms $V^a,\,\ \omega^{ab},\,\ \p_A,\,\ A^r$ become
superfield 1-forms whose physical meaning is given by the vielbein
$V^a$, the spin connection $\omega^{ab}$ and the gravitino
$\p_A$; eqs. (\ref{maucc}) then describe the vacuum configuration
in superspace whose bosonic subspace is $AdS_6$. On the ordinary
space-time, that is setting $\theta =0$ in the superfields
1-forms, the background vacuum fields have as $dx^{\mu}$
components the following expressions:
\begin{equation}
V^a_{\mu}= \delta^a_{\mu};\,\ \psi_{A\mu}=0; \,\
\left(\omega^{ab}_{\mu},\,\ A^r_{\mu}\right)= pure\,\ gauge.
\end{equation}

At this point, however, it is clear that equations (\ref{maucc})
do not describe the supersymmetric vacuum of the full $F(4)$
supergravity theory, because of the absence of the 2-form $B$ and
of the 1-form $A^0$ superfields, whose space-time restriction
coincides with the physical fields $B_{\mu\nu}$ and $A_{\mu}^0$
appearing in the supergravity multiplet.
 The recipe to have all the fields in
a single algebra is well known and consists in considering the
Free Differential Algebra (F.D.A.)\cite{bible} obtained from the
$F(4)$ M.C.E.'s by adding two more equations for the 2-form $B$
and for the 1-form $A^0$ (the 0-form fields $\chi_A$ and $\sigma$
do not appear in the algebra since they are set equal to
zero\footnote{Actually, the dilaton has to be set equal to a
constant value; however, by suitable redefinition we can set the
constant equal to zero.} in the vacuum). Using the tools explained
in reference \cite{bible} to construct F.D.A. containing forms of
higher degree (a 2-form in our case), it turns out that the only
consistent F.D.A. involving $B$ and $A^0$ is given by the
following extension of the (\ref{maucc}):
\begin{eqnarray}\label{dV}&&\mathcal{D}V^{a}-\frac{i}{2}\ \
\overline{\psi}_{A}\gamma_{a}\psi^A=0 \\
\label{dO}&&\mathcal{R}^{ab}+4m^{2}\ \
V^{a}V^{b}+m\overline{\psi}_{A}\gamma_{ab}\psi^A=0\\
 \label{dAr}&&dA^{r}+\frac{1}{2}\,\ g\,\
\epsilon^{rst}A_{s}A_{t}-i\,\ \overline{\psi}_{A}\psi_{B}\,\
\sigma^{rAB}=0\\
\label{dA}&&dA-mB-i\,\
\overline{\psi}_{A}\gamma_{7}\psi^A=0\\
\label{dB}&&dB+2\,\
\overline{\psi}_{A}\gamma_{7}\gamma_{a}\psi^AV^{a}=0\\
\label{dPsi}&&D\psi_{A}-im\gamma_{a}\psi_{A}V^{a}=0\end{eqnarray}
\noindent Equations, (\ref{dA}) and (\ref{dB}) were obtained by
imposing that together with the other Maurer Cartan equations they
satisfy the $d$-closure requirement. Actually the closure of
(\ref{dB}) relies on the 4-$\psi_A$'s Fierz identity (see
Appendix C)
\begin{equation}
\overline{\psi}_{A}\gamma_{7}\gamma_{a}\psi_{B}\epsilon^{AB}\overline{\psi}_{C}
\gamma^{a}\psi_{D}\epsilon^{CD}=0.
\end{equation}
The physical interesting feature of the F.D.A
(\ref{dV})-(\ref{dPsi}) describing the full supersymmetric vacuum
configuration is the appearance of the combination $dA^0-mB$ in
(\ref{dA}). When we go to the dynamical theory obtained by gauging
the F.D.A. out of the vacuum, the fields $A^0_{\mu}$ and
$B_{\mu\nu}$ will always appear in the same combination
$\partial_{[\mu}A^0_{\nu]}-mB_{\mu\nu}$. At the dynamical level
this implies, as noted by Romans \cite{rom}, an Higgs phenomenon
where the 2-form $B$ "eats" the 1-form $A^0$ and acquires a non
vanishing mass $m$.\\
In summary, we have shown that two of the main results of
\cite{rom}, namely  the existence of an $AdS$ supersymmetric
background only for $g=3m$ and  the Higgs-type mechanism by which
the field $B_{\mu\nu}$ becomes massive acquiring longitudinal
degrees of freedom in terms of the the vector $A^0_{\mu}$, are a
simple consequence of the algebraic structure of the F.D.A.
associated to the $F(4)$ supergroup written in terms of the
vacuum-superfields.\\
It is interesting to see what happens if one or both the
parameters $g$ and $m$ are zero. Setting $m=g=0$, one reduces the
$F(4)$ superalgebra to the $D=6\ \ N=(1,1)$ superalgebra existing
only in a super--Poincar\'e background; in this case the four-
vector $A^{\alpha}\equiv (A^0,A^r)$ transforms in the fundamental
of the $R$-symmetry group $SO(4)$ while the pseudo-Majorana
spinors $\psi_A,\chi_A$ can be decomposed in two chiral spinors,
as explained at the beginning of the section, in such a way that
all the resulting F.D.A. is invariant under
$SO(4)$.\\
Furthermore it is easy to see that no F.D.A  exists if either
$m=0$ , $g\neq 0$ or $m\neq 0$, $g= 0$, since the corresponding
equations in the F.D.A. do not close anymore under $d$-
differentiation. In other words, for a supersymmetric vacuum to
exist, the gauging of $SU(2)$, $g\neq 0$, must be necessarily
accompanied by the presence of the parameter $m$ which, as we
have seen, makes the closure of the supersymmetric algebra
consistent for $g=3m$.
%%%%%%%%%%%%%%%%%%%%%%%%%%%%%%%%%%%%%%%%%%%%%%%%%%%%%%%
%%%%%%%
\section{The ungauged theory}
\setcounter{equation}{0}
 In the previous section, we have fully
discussed the vacuum structure of the $F(4)$ supergravity which,
as we have shown, naturally admits an $AdS$ background when
$g=3m$. Our next task is to discuss the theory out of the vacuum,
that is to define appropriate curvatures for all the physical
fields and to retrieve the supersymmetry transformation laws and
the Lagrangian of the dynamical theory. The proper way to define
field strengths in the geometrical setting of superspace is to
introduce ``curvatures" defined as the deviation of the M.C.E.
from zero,
once the physical 1-forms are out of the vacuum.\\
However, as it happens in all supergravity theories, the dynamical
theory involves the presence of a non compact symmetry
($U$-duality). In our case  the presence of the dilaton, which is
set equal to zero in the vacuum, introduces a non compact
$O(1,1)$ symmetry, under which the four vectors $A^0$, $A^r$
transform non trivially. By suitable normalization of the dilaton
field we define the action of $O(1,1)$ as follows
 \be
(A^0,A^r)\longrightarrow e^{\s}(A^0,A^r). \ee
 The 2-form
$B$ is also charged under $O(1,1)$ transforming as
 \be
B\longrightarrow e^{-2\s}B. \ee
 This observation implies that when
we are out of the vacuum the two parameters $g$ and $m$ must scale
under $O(1,1)$ as follows
 \be  g\longrightarrow e^{-\s}g;\ \
m\longrightarrow e^{3\s}m\ee
 as it is evident from
equations (\ref{dAr}), (\ref{dA}). Following this prescription,
the gravitino curvature out of the vacuum would be defined as:
 \be
\rho_A=D\psi_A-ime^{-3\s}\gamma_{a}\psi_{A}V^{a}. \label{rho}
 \ee
 Note that the r.h.s. of \eq{rho} has no group--theoretical
 meaning in this case, because only in
the vacuum, where $\s =0$, it defines an $AdS$ covariant
derivative. For this reason, we try to build the theory starting
from a Poincar\'e invariant vacuum which is described by the
(\ref{maucc}) with $g=m=0$. Therefore we define the following set
of curvatures for the Poincar\'e  theory:
{\setlength\arraycolsep{1pt}\begin{eqnarray}
\label{Tg}&T^a&=\mathcal{D}V^a-\frac{i}{2}\ \
\overline{\psi}_A\gamma_a\psi^AV^a=0\\
\label{Lg}&R^{ab}&=\mathcal{R}^{ab}\\ \label{Hg} &H&=dB+2
e^{-2\sigma}\,\
\overline{\psi}_A\gamma_7\gamma_a\psi^AV^a\\
 \label{Fg}&F&=dA-ie^{\sigma}\,\
\overline{\psi}_A\gamma_7\psi^A\\
\label{Frg}&F^r&=dA^r-ie^{\sigma}\,\ \overline{\psi}_A\psi_B\,\
\sigma^{rAB}\\ \label{rog} &\rho_A&=D\psi_A\\
 \label{Rg}&R(\chi_A)&\equiv D\chi_A\\
 \label{sg}&R(\sigma)&\equiv
 d\sigma
\end{eqnarray}}
In reference \cite{noi}, starting from the super Poincar\'e
curvatures and the $SU(2)$ $R$-symmetry, we reproduced the results
of Romans \cite{rom} for pure supergravity theory. The coupling to
an arbitrary number of non abelian gauge matter multiplets was
worked out at the level of Bianchi identities. In this paper we
follow the more logical step of performing the coupling to an
arbitrary number of vector multiplets without gauging neither the
$R$-symmetry nor the vector multiplets. After the theory and its
Lagrangian have been constructed we turn to the problem of the
gauging and add the necessary new terms, proportional to the
gauge coupling constants, to the Lagrangian and to the
transformation laws.\\
Let us therefore first discuss the matter vector multiplets of
the theory; this is the only kind of supersymmetric matter in
$D=6$, $N=2$. The vector multiplet is given by:
\begin{equation}
(A_{\mu},\,\ \lambda_A,\,\ \phi^{\alpha})^I
\end{equation}
 where $\alpha=0,1,2,3$ and the index $I$ labels an arbitrary
number $n$ of such multiplets. As it is well known the $4n$
scalars parametrize the coset manifold $SO(4,n)/SO(4)\times
SO(n)$. Taking into account that the pure supergravity has a non
compact duality group $O(1,1)$ parametrized by $e^{\sigma}$, the
duality group of the matter coupled theory is
 \begin{equation}\label{dualitygroup}
  G=SO(4,n)\times O(1,1)
\end{equation}
To perform the matter coupling we follow the geometrical
procedure of introducing the coset representative $L^{\Lambda}_{\
\ \Sigma}$ of the matter coset manifold, where
$\Lambda,\Sigma,\dots=0, \dots, 4+n$; decomposing the $SO(4,n)$
indices with respect to $H=SO(4)\times SO(n)$ we have:
\begin{equation}
L^{\Lambda}_{\ \ \Sigma}=(L^{\Lambda}_{\ \ \alpha},L^{\Lambda}_{\
\ I})
\end{equation}
\noindent where $\alpha=0,1,2,3$ and $I=4,\dots ,4+n$.
Furthermore, since we are going to gauge the $SU(2)$ diagonal
subgroup of $SO(4)$ as in pure supergravity, we will also
decompose $L^{\Lambda}_{\ \ \alpha}$ as
\begin{equation}
L^{\Lambda}_{\ \ \alpha}=(L^{\Lambda}_{\ \ 0}, L^{\Lambda}_{\ \
r}), \quad \mbox{with  } r=1,2,3.
\end{equation}
The $4+n$ gravitational and matter vectors transform in the
fundamental of $SO(4,n)$ so that the superspace curvatures of the
vectors will be now labeled by the index $\Lambda$. Furthermore
the covariant derivatives acting on the spinor fields will now
contain also the composite connections of the $SO(4,n)$ duality
group.
\par
 Let us introduce the left-invariant 1-form of $SO(4,n)$
\begin{equation}
\Omega^{\Lambda}_{\ \ \Sigma}=(L^{\Lambda}_{\ \ \Pi})^{-1}
dL^{\Pi}_{\ \ \Sigma}
\end{equation}
 \noindent satisfying the Maurer-Cartan
equation
\begin{equation}
d\Omega^{\Lambda}_{\ \ \Sigma}+\Omega^{\Lambda}_{\ \
\Pi}\land\Omega^{\Pi}_{\ \ \Sigma}=0
\end{equation}
\noindent By appropriate decomposition of the indices, we find:
\begin{eqnarray}
\label{1}&&R^r_{\,\ s}=-P^{r}_{\ \ I}\land P^I_{\ \ s}\\
\label{2}&&R^r_{\,\ 0}=-P^{r}_{\ \ I}\land P^I_{\ \ 0}\\
\label{3}&&R^I_{\,\ J}=-P^I_{\ \ r}\land P^r_{\ \ J}-P^I_{\ \
0}\land P^0_{\ \ J}\\ \label{4}&&\nabla P^I_{\,\ r}=0\\
\label{5}&&\nabla P^I_{\,\ 0}=0
\end{eqnarray}
\noindent where
\begin{eqnarray}
&&R^{rs}\equiv d\Omega^r_{\ \ s}+\Omega^{r}_{\ \
t}\land\Omega^t_{\ \ s}+\Omega^{r}_{\ \ 0}\land\Omega^0_{\ \ s}\\
&&R^{r0}\equiv d\Omega^r_{\ \ 0}+\Omega^{r}_{\ \
t}\land\Omega^t_{\ \ 0}\\ &&R^{IJ}\equiv d\Omega^I_{\ \
J}+\Omega^{I}_{\ \ K}\land\Omega^K_{\ \ J}
\end{eqnarray}
\noindent and we have set
\begin{displaymath}
P^I_{\alpha}=\left\{ \begin{array}{rr}P^I_{\,\ 0}\equiv
\Omega^{I}_{\ \ 0}\\ P^I_{\,\ r}\equiv \Omega^{I}_{\ \
r}\end{array}\right.
\end{displaymath}
\noindent Note that $P^I_0$, $P^I_r$ are the vielbeins of the
coset, while $(\Omega^{rs},\,\ \Omega^{r0})$, $(R^{rs},\,\
R^{ro})$ are respectively the connections and the curvatures of
$SO(4)$ decomposed with respect to the diagonal subgroup
$SU(2)\subset SO(4)$.\\ In terms of the previous definitions, the
ungauged superspace curvatures of the matter coupled theory,
 are now
modified, with respect to eqs. (\ref{Tg}) - (\ref{sg}),
 in two aspects:  first, in the definition of the
superspace vector field strengths, there appear, besides the
$O(1,1)$ representative $e^{\s}$, also the coset representatives
of the $G/H$ $\s$-model, which intertwine between the $R$-symmetry
indices $A,\,\ B,\dots$ of the gravitinos and the indices $\L,\,\
\Sigma,\dots$ of the $4+n$-dimensional $G$ representation;
secondly, the definitions of the fermion curvatures are modified
by the presence of the $SU(2)$ connection acting on gravitinos and
dilatinos, and the $SU(2)$ and $SO(n)$ connection on the gauginos.
Therefore the ungauged superspace curvatures of the matter coupled
theory are now given by:
{\setlength\arraycolsep{1pt}\begin{eqnarray}
&T^{A}&=\mathcal{D}V^{a}-\frac{i}{2}\ \
\overline{\psi}_{A}\gamma_{a}\psi^A V^{a}=0\nonumber\\
&R^{ab}&=\mathcal{R}^{ab}\nonumber\\
&H&=dB+2 e^{-2\sigma}\,\
\overline{\psi}_{A}\gamma_{7}\gamma_{a}\psi^AV^{a}\nonumber\\
&F^{\Lambda}&=\mathcal{F}^{\Lambda}-ie^{\sigma}L^{\Lambda}_0\epsilon^{AB}
\overline{\psi}_{A}\gamma_{7}\psi_{B}-ie^{\sigma}L^{\Lambda}_r\sigma^{rAB}
\overline{\psi}_{A}\psi_{B}\nonumber\\
&\rho_{A}&=\mathcal{D}\psi_{A}-\frac{i}{2}\sigma_{rAB}(-\frac{1}{2}
\epsilon^{rst}\Omega_{st}-i\gamma_7\Omega_{r0})\psi^B\nonumber\\
&D\chi_{A}&=\mathcal{D}\chi_{A}-\frac{i}{2}\sigma_{rAB}
(-\frac{1}{2}\epsilon^{rst}\Omega_{st}-i\gamma_7\Omega_{r0})\chi^B\nonumber\\
&R(\sigma)&=d\sigma\nonumber\\
&\nabla\lambda_{IA}&=\mathcal{D}\lambda_{IA}-\frac{i}{2}\sigma_{rAB}
(-\frac{1}{2}\epsilon^{rst}\Omega_{st}-i\gamma_7\Omega_{r0})\lambda_I^B
+\Omega_I^{\ \ J}\l_{JA}\nonumber\\
&R^{I}_0(\phi)&\equiv P^{I}_0\nonumber\\
&R^{I}_r(\phi)&\equiv P^{I}_r \label{curve}
\end{eqnarray}}
\noindent where the last two equations define the "curvatures" of
the matter scalar fields $\phi^i$ as the vielbein of the coset:
\begin{equation}
P^{I}_0\equiv P^{I}_{0 i}d\phi^i\ \ \ \ P^{I}_r\equiv P^{I}_{r
i}d\phi^i
\end{equation}
\noindent where $i$ runs over the $4n$ values of the coset
vielbein world-components.\\
By straightforward computation we obtain the Bianchi identities:
\begin{eqnarray}
\label{xtors}&&R^{ab} V_{b}-i
\overline{\psi}_{A}\gamma^{a}\rho_{B}\epsilon^{AB}=0\\
&&\mathcal {D}R^{ab}=0\\
\label{xdH}&&dH+4e^{-2\sigma}d\sigma\,\
\overline{\psi}_{A}\gamma_{7}\gamma_{a}\psi_{B}\epsilon^{AB}V^{a}+4e^{-2\sigma}
\overline{\psi}_{A}\gamma_{7}\gamma_{a}\rho_{B}\epsilon^{AB}V^{a}=0\\
 \label{xdF}&&DF^{\Lambda}+id\sigma
e^{\sigma}\overline{\psi}_{A}\gamma_{7}\psi_{B}L^{\Lambda}_{[AB]}+id\sigma
e^{\sigma}\overline{\psi}_{A}\psi_{B}L^{\Lambda}_{(AB)}\nonumber\\
&&-2ie^{\sigma}
\overline{\psi}_{A}\gamma_{7}\rho_{B}L^{\Lambda}_{[AB]}
-2ie^{\sigma}
\overline{\psi}_{A}\rho_{B}L^{\Lambda}_{(AB)}+ie^{\s}L^{\L}_{\ \
I}\pb^A\p^BP^I_{(AB)}+\nonumber\\
&&+ie^{\s}L^{\L}_{\ \ I}\pb^A\g_7\p^BP^I_{[AB]}=0\\
\label{xdro}&&D\rho_{A}+\frac{1}{4}R^{ab}\gamma_{ab}\psi_{A}-\frac{i}{2}\sigma^{r}_{AB}(\frac{1}{2}\epsilon^{rst}\mathcal{R}^{st}
+i\gamma_{7}\mathcal{R}_{r0})\psi^{B}=0\\
\label{xdR}&&D^2\chi_{A}+\frac{1}{4}R^{ab}\gamma_{ab}\chi_{A}-\frac{i}{2}\sigma^{r}_{AB}(\frac{1}{2}\epsilon^{rst}\mathcal{R}^{st}
+i\gamma_{7}\mathcal{R}_{r0})\chi^{B}=0\\
&&d^{2}\sigma=0\\
\label{xd2l}&&D^{2}\lambda^{I}_{A}+\frac{1}{4}R^{ab}\gamma_{ab}\lambda^{I}_{A}
-\frac{i}{2}\sigma^{r}_{AB}(\frac{1}{2}\epsilon^{rst}\mathcal{R}^{st}+i\gamma_{7}\mathcal{R}_{r0})\lambda^{IB}
-\mathcal{R}^{I}_{J}\lambda^{J}_{A}=0\\
\label{xdP}&&DP^{I}_{AB}=0
\end{eqnarray}
 where $P^I_{AB}=P^I_0\epsilon_{AB}+P^I_r\s^r_{AB}$.\\
The solution of the Bianchi identities is a quite non trivial
task, especially when examined in the sector involving the 3
gravitino 1-forms, because, as we will show in Appendix A and C,
we need terms of the form $\p\p\chi$ in the gravitino curvature
superspace parametrization in  order to have a consistent
solution of both the fermionic and bosonic Bianchi identities.
This in turn implies a full mastering of all the Fierz identities
connecting different 3-$\p$ expressions. In Appendix A we give a
short account of the various techniques used in order to solve
our problem, together with the solution of Bianchi identities
in superspace.\\
Here we limit ourselves to present the solution in terms of the
space-time transformation laws of the physical fields which, as
is well known, can be immediately written down once the
parametrizations of the supercurvatures in superspace are found.
We have: {\setlength\arraycolsep{1pt}\begin{eqnarray} \label
{t1}&\delta
V^{a}_{\mu}&=-i\overline{\psi}_{A\mu}\gamma^{a}\varepsilon^A\\
\label {t2} &\delta B_{\mu\nu}&=4\ii e^{-2\sigma}
\overline{\chi}_{A}\gamma_{7}\gamma_{\mu\nu}\varepsilon^A
-4e^{-2\sigma}\overline{\varepsilon}_A\gamma_7\gamma_{[\mu}\psi_{\nu]}^A\\
 \label {t3}&\delta A^{\Lambda}_{\mu}&=2 e^{\sigma}
 \overline{\varepsilon}^{A}\gamma_{7}\gamma_{\mu}\chi^BL^{\Lambda}_0
 \epsilon_{AB}+2e^{\sigma}\overline{\varepsilon}^{A}\gamma_{\mu}\chi^{B}
 L^{\Lambda r}\sigma_{rAB}-e^{\sigma}L_{\Lambda
I}\overline{\varepsilon}^{A}\gamma_{\mu}\lambda^{IB}\epsilon_{AB}+\nonumber\\
&&+2ie^{\sigma}L^{\Lambda}_0\overline{\varepsilon}_A\gamma^7\psi_B\epsilon^{AB}+
2ie^{\sigma}L^{\Lambda r}\sigma_{r}^{AB}\overline{\varepsilon}_A\psi_B\\
\label{qui}&\delta\psi_{A\mu}&=\mathcal{D}_{\mu}\varepsilon_A+\frac{1}{16}
e^{-\sigma}[T_{[AB]\nu\lambda}\gamma_{7}-T_{(AB)\nu\lambda}](\gamma_{\mu}^{\,\
\nu\lambda}-6\delta_{\mu}^{\nu}\gamma^{\lambda})
\varepsilon^{B}+\nonumber \\
 &&+\frac{i}{32}e^{2\sigma} H_{\nu\lambda\rho}
\gamma_{7}(\gamma_{\mu}^{\,\ \nu\lambda\rho}-3\delta_{\mu}^{\nu}
\gamma^{\lambda\rho})\varepsilon_{A}+\frac{1}{2}\varepsilon_{A}\overline{\chi}^{C}\psi_{C\mu}+\nonumber\\
&&+\frac{1}{2}\gamma_{7}\varepsilon_{A}\overline{\chi}^{C}\gamma^{7}\psi_{C\mu}-
\gamma_{\nu}\varepsilon_{A}\overline{\chi}^{C}\gamma^{\nu}\psi_{C\mu}+\gamma_{7}
\gamma_{\nu}\varepsilon_{A}\overline{\chi}^{C}\gamma^{7}\gamma^{\nu}\psi_{C\mu}+\nonumber\\
&&-\frac{1}{4}\gamma_{\nu\lambda}\varepsilon_{A}\overline{\chi}^{C}\gamma^{\nu\lambda}\psi_{C\mu}-
\frac{1}{4}\gamma_{7}\gamma_{\nu\lambda}\varepsilon_{A}\overline{\chi}^{C}\gamma^{7}\gamma^{\nu\lambda}\psi_{C\mu}\\
\label{quo}&\delta\chi_{A}&=\frac{i}{2}
\gamma^{\mu}\partial_{\mu}\sigma \varepsilon_{A}+
\frac{i}{16}e^{-\sigma}[T_{[AB]\mu\nu}\gamma_{7}+T_{(AB)\mu\nu}]\gamma^{\mu\nu}\varepsilon^{B}+\nonumber\\
&&+\frac{1}{32}e^{2\sigma}
H_{\mu\nu\lambda}\gamma_{7}\gamma^{\mu\nu\lambda}\varepsilon_{A}\\
\label {t4}&\delta\sigma&=\overline{\chi}_{A}\varepsilon^A\\
\label{qua}&\delta\lambda^{IA}&=-iP^I_{ri}\sigma^{rAB}\partial_{\mu}\phi^{i}\gamma^{\mu}\varepsilon_{B}+iP^I_{0i}\epsilon^{AB}\partial_{\mu}\phi^{i}\gamma^{7}\gamma^{\mu}\varepsilon_{B}+\frac{i}{2}e^{-\sigma}T^{I}_{\mu\nu}\gamma^{\mu\nu}\varepsilon^{A}\\
\label {t5}&P^{I}_{0i}\delta\phi^i&=\frac{1}{2}\overline{\lambda}^{I}_{A}\gamma_{7}\varepsilon^A\\
\label {t6}
&P^{I}_{ri}\delta\phi^i&=\frac{1}{2}\overline{\lambda}^{I}_{A}\varepsilon_{B}\sigma_r^{AB}
\end{eqnarray}}
\noindent where we have introduced the "dressed" vector field
strengths
\begin{eqnarray}
&&T_{[AB]\mu\nu}\equiv\epsilon_{AB}L^{-1}_{0\Lambda
}F^{\Lambda}_{\mu\nu}\\
&&T_{(AB)\mu\nu}\equiv\sigma^r_{AB}L^{-1}_{r\Lambda}F^{\Lambda}_{\mu\nu}\\
&&T_{I\mu\nu}\equiv L^{-1}_{I\Lambda }F^{\Lambda}_{\mu\nu}
\end{eqnarray}
\noindent and we have omitted in the transformation laws of the fermions the
three-fermions terms of the form
$(\chi\chi\varepsilon)$, $(\lambda\lambda\varepsilon)$, $(\lambda\chi\varepsilon)$.\\
Instead, we have included  the extra terms in the gravitino
transformation law of the form $\psi \chi \varepsilon$ which, like
all the three-fermion terms, were not computed in reference
\cite{rom}. However these terms correspond to terms $\psi \psi
\chi$ in the superspace curvature $D\psi_A$ which, as we discuss
in Appendix, are quite essential to verify the consistency of the
Bianchi identities; therefore they have an important meaning for
the consistence of the theory and this is the reason why we have
explicitly quoted them. This is to be contrasted with other
three-fermion terms of the form $\chi \chi \varepsilon$,
$(\lambda\lambda\varepsilon)$, $(\lambda\chi\varepsilon)$  on
space-time (that is ($\chi \chi \psi$), $(\lambda\lambda\p)$,
$(\lambda\chi\p)$ in superspace), which we have not included in
the transformation law of the fermions. Indeed their explicit form
can always be  found from the Bianchi identities once the
consistency in the higher fermionic sectors has been verified, so
that they are not on the same status. Since their explicit
computation is very cumbersome they have not been computed.

An important property of the solution presented is the fact that,
in the ungauged theory, no supersymmetric $AdS$ background
exists, as it was expected from the discussion given in Section 2,
while the Poincar\'e supersymmetric vacuum does exist. Indeed if
we go in the vacuum, where all the field strengths of the bosonic
fields and the fermionic fields are zero , and the scalar $\s$
takes an arbitrary constant value (which we set equal to zero),
one easily derives that under supersymmetry eqs. \eq{qui},
\eq{quo}, \eq{qua} reduce to:
\begin{eqnarray}
&&\delta\psi_{A\mu}=\mathcal{D}_{\mu}\varepsilon_A\\
&&\delta\chi_A=0\\
&&\delta\lambda^{IA}=0\\
\end{eqnarray}
which proves our statement. In the next section, we will gauge
the theory and we will see that in presence of  gauging we find a
more general solution of the Bianchi identities containing a new
parameter $m$, such that in the
vacuum a supersymmetric $AdS_6$ background can be retrieved.\\
We observe that the solutions of the Bianchi identities also
imply the equations of motion of the physical fields and
therefore one can reconstruct in principle the space-time
Lagrangian.  Nevertheless, in general, it is simpler to construct
the Lagrangian explicitly. In order to do that, we use a
superspace geometric approach, called ``rheonomic'' \cite{bible},
for the construction of a Lagrangian in superspace, which turns
out to be greatly simplified once the Bianchi identities have been
solved. Its construction is sketched in Appendix B. Here we just
quote the space-time Lagrangian which is obtained from the
rheonomic one by restricting all the fields to their space-time
values (see Appendix B). The final expression is the following:
\be\mathcal{A}=\int\mathcal{L}^{(D=6,N=2)}_{(ungauged)}\sqrt{-g}\,\
d^6x \ee
 where:
  \be  \mathcal{L}^{(D=6,N=2)}_{(ungauged)}\equiv
\mathcal{L}_{\mbox{\tiny{kin}}}
+\mathcal{L}_{\mbox{\tiny{Pauli}}}+
\mathcal{L}_{\mbox{\tiny{Chern--Simons}}}
+\mathcal{L}_{\mbox{\tiny{4 fermions}}} \label{lagrungauged} \ee
and:
%%%%%%%%%%%%%%%%%%%%%%%%%%%%%%%%%%%%%%%%%%%%%%%%%%%%%
%%%%%%%%%lagrangiana cinetica %%%%%%%%%%%%%%%%%%%%%
%%%%%%%%%%%%%%%%%%%%%%%%%%%%%%%%%%%%%%%%%%%%%%%%%%%%
 \ba \mathcal{L}_{\mbox{\tiny{kin}}}&=&
-\frac{1}{4}\mathcal{R}-\frac{1}{8}e^{2\sigma}\mathcal{N}_{\Lambda\Sigma}
{\mathcal{F}}^{\Lambda}_{\mu\nu}{\mathcal{F}}^{\Sigma\mu\nu}
+\frac{3}{64}e^{4\s}H_{\mu\nu\rho}H^{\mu\nu\rho}+\nonumber\\
&&+\frac{i}{2}\overline{\psi}_{A\mu}\gamma^{\mu\nu\rho}
D_{\nu}\psi^{A}_{\rho} -
2i\overline{\chi}_A\gamma^{\mu}D_{\mu}\chi^A+
\frac{i}{8}\overline{\lambda}^I_A\gamma^{\mu}D_{\mu}\lambda^A_I+\nonumber\\
&&+
\partial^{\mu}\sigma\partial_{\mu}\sigma  -\frac{1}{4}
\left(P^{I0}_{i} P_{I0 j} +P^{Ir}_{i}P_{Ir  j}\right)
\partial^{\mu}\phi^i\partial_{\mu}\phi^{j}
 \, ;\ea
%%%%%%%%%%%%%%%%%%%%%%%%%%%%%%%%%%%%%%%%%%%%%%%%%%%%%
%%%%%%%%%lagrangiana di pauli %%%%%%%%%%%%%%%%%%%%%
%%%%%%%%%%%%%%%%%%%%%%%%%%%%%%%%%%%%%%%%%%%%%%%%%%%%
\ba{\mathcal{L}}_{\mbox{\tiny{Pauli}}}&=&-2\partial_{\mu}\s
\bar\chi_A\g^{\nu}\g^{\mu}\p^A_{\nu}
+\frac{1}{4}P^{I0}_{i}\partial^{\mu}\phi^i
\bar\lambda_{IA}\g_7\g^{\nu}\g^{\mu}\p^A_{\nu}
+\frac{1}{4}P^{I(AB)}_{i}\partial^{\mu}\phi^i
\bar\lambda_{IA}\g^{\nu}\g^{\mu}\p_{B\nu}+\nonumber\\
&+\!&e^{-\s}\mathcal{N}_{\L\Sigma}{\mathcal{F}}^{\L}_{\mu\nu}
\Big\{\frac{i}{8}L^{\Sigma}_0\pb_A^{\rho}
\g_7\left(\g^{\mu\nu}_{\phantom{\mu\nu}\rho\s}+2\delta^{\mu\nu}_{\rho\s}\right)\p^{A\s}+
\frac{i}{8}L^{\Sigma}_{(AB)}\pb^{A\rho}
\left(\g^{\mu\nu}_{\phantom{\mu\nu}\rho\s}+2\delta^{\mu\nu}_{\rho\s}\right)\p^{B\s}\nonumber\\
&-&\frac{1}{4}L^{\Sigma}_0
\pb_{A\rho}\g_7\g^{\mu\nu}\g^{\rho}\chi^A
-\frac{1}{4}L^{\Sigma}_{(AB)}\pb_{\rho}^A\g^{\mu\nu}\g^{\rho}\chi^B
-\frac{1}{8}L^{\Sigma}_I
\bar\lambda^I_A\g^{\mu\nu}\g^{\rho}\psi^A_{\rho}+\nonumber\\
&+&\frac{1}{2}L^{\Sigma}_0
\left(\frac{\ii}{2}\bar\chi_{A}\g_7\g^{\mu\nu}\chi^A+\frac{\ii}{16}\bar\lambda_{IA}\g_7\g^{\mu\nu}\l^{IA}\right)
+ \nonumber\\
&+&\frac{1}{2}L^{\Sigma}_{(AB)}
\left(-\frac{\ii}{2}\bar\chi^{A}\g^{\mu\nu}\chi^B+\frac{\ii}{16}\bar\lambda_{I}^{A}\g^{\mu\nu}\l^{IB}\right)
\Big\}+\nonumber\\
&+&\frac{3}{8}e^{2\s}H_{\mu\nu\rho}\Big\{\frac12\pb_A^{\l}\g_7
\left(\delta^{\mu\nu}_{\l\s}\g^{\rho}-16\g^{\mu\nu\rho}_{\phantom{\mu\nu\rho}\l\s}\right)\p^{A\s}
+\frac{i}{3}\pb_{A\s}\g_7\g^{\s}\g^{\mu\nu\rho}\chi^A+\nonumber\\
&+&\frac{1}{3}\bar\chi_A\g_7\g^{\mu\nu\rho}\chi^A\Big\};\ea
%%%%%%%%%%%%%%%%%%%%%%%%%%%%%%%%%%%%%%%%%%%%%%%%%%%%%
%%%%%%%%%lagrangiana di chern-simons %%%%%%%%%%%%%%%
%%%%%%%%%%%%%%%%%%%%%%%%%%%%%%%%%%%%%%%%%%%%%%%%%%%%
\be \mathcal{L}_{\mbox{\tiny{Chern--Simons}}}=
-\frac{1}{64}\epsilon^{\mu\nu\rho\s\l\tau}B_{\mu\nu}
\eta_{\L\Sigma}{\mathcal{F}}^{\L}_{\rho\s}{\mathcal{F}}^{\Sigma}_{\l\tau}
\ee
where we have defined \cite{adf}:
\begin{equation}
{\cal N}_{\Lambda\Sigma} =(L^{-1})_{AB | \Lambda}
(L^{-1})^{AB}_{\phantom{AB}\Sigma} -(L^{-1})_{I | \Lambda}
(L^{-1})^{I}_{\phantom{I}\Sigma} .\label{kin}
\end{equation}\\
%%%%%%%%%%%%%%%%%%%%%%%%%%%%%%%%%%%%%%%%%%%%%%%%%%%%%%
%%%%%%%%  IL GAUGING  %%%%%%%%%%%%%%%%%%%%%%%%%%%%%%%
%%%%%%%%%%%%%%%%%%%%%%%%%%%%%%%%%%%%%%%%%%%%%%%%%%%%%%
\section{The gauging}
\setcounter{equation}{0}
The next problem we have to cope with is
the gauging of the matter coupled theory and the determination of
the scalar
potential.\\
Let us first consider the ordinary gauging, with $m=0$, which, as
usual, will imply the presence of new terms proportional to the
coupling constants in the supersymmetry transformation laws of
the fermion fields.\\
Our aim is to gauge a compact subgroup of $SO(4,n)$. Since in any
case we may gauge only the diagonal subgroup $SU(2)\subset
SO(4)\subset SO(4)\otimes SO(n)$, the maximal gauging is given by
$SU(2)\otimes\mathcal{G}$ where $\mathcal{G}$ is a $n$-dimensional
subgroup of $SO(n)$. According to a well known procedure, we
modify the definition of the left invariant 1-form $L^{-1}dL$ by
replacing the ordinary differential with the
$SU(2)\otimes\mathcal{G}$ covariant differential as follows:
\begin{equation}
\label{nabla}\nabla L^{\Lambda}_{\ \ \Sigma}=d L^{\Lambda}_{\ \
\Sigma}-f_{\Gamma\ \ \Pi}^{\,\ \Lambda} A^{\Gamma} L^{\Pi}_{\ \
\Sigma}
\end{equation}
\noindent where $f^{\Lambda}_{\ \ \Pi\Gamma}$ are the structure
constants of $SU(2)\otimes\mathcal{G}$, $SU(2)$ being the
diagonal subgroup of $SU(2)_L \otimes SU(2)_R \simeq SO(4)$. More
explicitly, denoting with $\epsilon^{rst}$ and
$\mathcal{C}^{IJK}$ the structure constants of the two factors
$SU(2)$ and $\mathcal{G}$, equation (\ref{nabla}) splits as
follows:
\begin{eqnarray}
&&\nabla L^{0}_{\ \ \Sigma}=d L^{\Lambda}_{\ \ \Sigma}\\
&&\nabla L^{r}_{\ \ \Sigma}=d L^{r}_{\ \ \Sigma}-g\epsilon^{\,\
r}_{t\
\ s} A^{t} L^{s}_{\ \ \Sigma}\\
&&\nabla L^{I}_{\ \ \Sigma}=d L^{I}_{\ \
\Sigma}-g'\mathcal{C}^{\,\ I}_{K\ \ J} A^{K} L^{J}_{\ \ \Sigma}
\end{eqnarray}
\noindent Setting $\widehat{\Omega}=L^{-1}\nabla L$, one easily
obtains the gauged Maurer-Cartan equations:
\begin{equation}\label{mc}d\widehat{\Omega}^{\Lambda}_{\ \
\Sigma}+\widehat{\Omega}^{\Lambda}_{\ \
\Pi}\land\widehat{\Omega}^{\Pi}_{\ \
\Sigma}=(L^{-1}\mathcal{F}L)^{\Lambda}_{\ \ \Sigma}
\end{equation}
\noindent where $\cF\equiv\cF^{\Lambda}T_{\Lambda}$, $T_{\Lambda}$ being the generators of $SU(2)\otimes\cG$.\\
After gauging, the same decomposition as in eqs. (\ref{1})
-(\ref{5}) now gives:
\begin{eqnarray}
\label{11}&&\widehat{R}^r_{\,\ s}=R^r_{\,\ s}+(L^{-1}\mathcal{F}L)^{r}_{\ \ s} = - P^r_I \wedge P^I_s +
(L^{-1}\mathcal{F}L)^{r}_{\ \ s} \\
\label{12}&&\widehat{R}^r_{\,\ 0}=R^r_{\,\
}+(L^{-1}\mathcal{F}L)^{r}_{\ \ 0} = - P^r_I \wedge P^I_0 +
(L^{-1}\mathcal{F}L)^{r}_{\ \ 0}\\
\label{13}&&\widehat{R}^I_{\,\ J}=R^I_{\,\
J}+(L^{-1}\mathcal{F}L)^{I}_{\ \ J} =  -P_r^I \wedge P_J^r -P_0^I
\wedge P_J^0 +
(L^{-1}\mathcal{F}L)^{I}_{\ \ J}\\
\label{14}&&\nabla \widehat{P}^I_{\,\ r}=(L^{-1}\mathcal{F}L)^{I}_{\ \ r}\\
\label{15}&&\nabla \widehat{P}^I_{\,\
0}=(L^{-1}\mathcal{F}L)^{I}_{\ \ 0}
\end{eqnarray}
where: \begin{eqnarray} \widehat{P}^I_{\
0}&=&\left(L^{-1}\right)^I_{\ \L} \nabla L^\L_{\,\ 0}\nonumber\\
\widehat{P}^I_{\,\ r}&=&\left(L^{-1}\right)^I_{\,\ \L} \nabla
L^\L_{\,\ r}. \label{scalviel}
\end{eqnarray}

Because of the presence of the gauged terms in the coset
curvatures, the new Bianchi Identities (whose explicit form is
given in Appendix A) are not satisfied by the old superspace
curvatures. Therefore we obtain a different solution for the
``curvatures'' which, in space--time language, amounts to
different transformation laws. However, the new solution for the
curvatures and hence the new transformation laws can be obtained
from the old ones by performing in  \eq{t1}--\eq{t6} the
following modifications:
\begin{enumerate}
\item{The vector field--strengths $F^\Lambda_{\mu\nu}$ are now non
abelian.}
\item{For the vielbein of the scalar manifold, we perform the
replacement: $(P^I_{0 i} , P^I_{r i}) \to (\widehat{P}^I_{0 i} ,
\widehat{P}^I_{r i})$, according to equation \eq{scalviel}.}
\item{This is the most important modification: the transformation
rules of the Fermi fields require extra terms proportional to the
gauge coupling constants, called ``fermionic shifts''. In
particular, if we denote by ,
$\widehat{\delta\psi}_{A\mu}^{(old)}$,
$\widehat{\delta\chi}_A^{(old)}$ and $\widehat{\delta\lambda}^{I
(old)}_A$ the transformation laws (\ref{qui}), (\ref{quo}),
(\ref{qua}) modified according to items 1. and 2. we may write:
\begin{eqnarray}
\label{del1}&&\delta\psi_{A\mu}=\widehat{\delta\psi}_{A\mu}^{(old)}+S_{AB}(g,g')\gamma_{\mu}\varepsilon^B\\
\label{del2}&&\delta\chi_A=\widehat{\delta\chi}_A^{(old)}+N_{AB}(g,g')\varepsilon^B\\
\label{del3}&&\delta\lambda_A^I=\widehat{\delta\lambda}^{I
(old)}_A+M^I_{AB}(g,g')\varepsilon^B
\end{eqnarray}
} \end{enumerate}
 \noindent Working out the Bianchi
identities, one fixes the explicit form of the fermionic shifts
which turn out to be
\begin{eqnarray}
\label{S}&&S_{AB}^{(g,g')}=\frac{i}{24}Ae^{\sigma}\epsilon_{AB}-\frac{i}{8}B_t\gamma^7\sigma^t_{AB}\\
\label{N}&&N_{AB}^{(g,g')}=\frac{1}{24}Ae^{\sigma}\epsilon_{AB}+\frac{1}{8}B_t\gamma^7\sigma^t_{AB}\\
\label{M}&&M^{I(g,g')}_{AB}=(-C^I_t+2i\gamma^7D^I_t)\sigma^t_{AB}
\end{eqnarray}
\noindent where
\begin{eqnarray}\label{AA}&&A=\epsilon^{rst}K_{rst}\\  \label{BB}&&B^i=\epsilon^{ijk}K_{jk0}\\ \label{CC}&&C_I^t=\epsilon^{trs}K_{rIs}\\ \label{DD}&&D_{It}=K_{0It}\end{eqnarray}
\noindent and the threefold completely antisymmetric tensors
$K's$ are the so called "boosted structure constants" given
explicitly by:
\begin{eqnarray}
&&K_{rst}=g\epsilon_{lmn}L^l_{\,\ r}(L^{-1})^{\,\ m}_sL^n_{\,\
t}+g'\mathcal{C}_{IJK}L^I_{\,\ r}(L^{-1})^{\,\ J}_sL^K_{\,\ t}\\
&&K_{rs0}=g\epsilon_{lmn}L^l_{\,\ r}(L^{-1})^{\,\ m}_sL^n_{\,\
0}+g'\mathcal{C}_{IJK}L^I_{\,\ r}(L^{-1})^{\,\ J}_sL^K_{\,\ 0}\\
&&K_{rIt}=g\epsilon_{lmn}L^l_{\,\ r}(L^{-1})^{\,\ m}_IL^n_{\,\
t}+g'\mathcal{C}_{LJK}L^L_{\,\ r}(L^{-1})^{\,\ J}_IL^K_{\,\ t}\\
&&K_{0It}=g\epsilon_{lmn}L^l_{\,\ 0}(L^{-1})^{\,\ m}_IL^n_{\,\
t}+g'\mathcal{C}_{LJK}L^L_{\,\ 0}(L^{-1})^{\,\ J}_IL^K_{\,\ t}
\end{eqnarray}

However, this is not the most general solution of the Bianchi
identities. Indeed, even in absence of scalar matter fields, there
is a more general solution of the Bianchi identities which
involve a new parameter $m$ which behaves like a second
``gauging'' since it only affects the transformation laws of
Fermi fields with suitable shifts proportional to $m$ in the
transformation laws of the fermions, provided we also redefine the
singlet vector field-strength as: \be  F \to \hat F \equiv F -mB
\ee These $m$-shifts in the pure gravitational case are given by:
\begin{eqnarray}
S_{AB}^{(m)}&=&\frac{\ii}{4} m e^{-3\sigma}\gamma_7\epsilon_{AB}\\
N_{AB}^{(m)}&=& -\frac{3}{4}me^{-3\sigma}\epsilon_{AB}\\
\end{eqnarray}
which, in presence of matter multiplets, generalize to:
\begin{eqnarray}
\label{mS}
S_{AB}^{(m)}&=&\frac{\ii}{4}me^{-3\sigma}(L^{-1})_{00}\epsilon_{AB}
+\frac{i}{4}me^{-3\sigma}(L^{-1})_{i0}\gamma_7\sigma^t_{AB}\\
\label{mN}
N_{AB}^{(m)}&=&-\frac{3}{4}me^{-3\sigma}(L^{-1})_{00}\epsilon_{AB}
+\frac{3}{4}me^{-3\sigma}(L^{-1})_{i0}\gamma_7\sigma^t_{AB}\\
\label{mM}M^{I(m)}_{AB}&=&-2me^{-3\sigma}(L^{-1})^I_{\ \
0}\epsilon_{AB}
\end{eqnarray}
Hence, we may have two different theories either gauging
$SU(2)\times \cG$, with shifts given by eqs \eq{S}--\eq{M}, or
performing the second gauging proportional to $m$ with shifts
given by eqs. \eq{mS}--\eq{mM}. Neither of these two theories
have an invariant supersymmetric Poincar\'e or Anti de Sitter
background what is in agreement with the discussion of the F.D.A.
given at the end of section 2. Actually, these two gaugings do not
interfere and we may perform both at once obtaining in this way
the following general form for the fermionic shifts:
\begin{eqnarray}
S_{AB}^{(g,g',m)}&=&\!\frac{\ii}{24}[Ae^{\sigma}\! +\!
6me^{-3\sigma}(L^{-1})_{00}]\epsilon_{AB}\! -\!
\frac{i}{8}[B_te^{\sigma}-2me^{-3\sigma}(L^{-1})_{i0}]\gamma^7\sigma^t_{AB}\label{mgS}
\\
N_{AB}^{(g,g',m)}&=&\!\frac{1}{24}[Ae^{\sigma}\! -\!
18me^{-3\sigma}(L^{-1})_{00}]\epsilon_{AB}\! +\!
\frac{1}{8}[B_te^{\sigma}\! +\! 6me^{-3\sigma}(L^{-1})_{i0}]\gamma^7\sigma^t_{AB}\label{mgN}\\
\label {fine}
M^{I(g,g',m)}_{AB}&=&\!(-C^I_t+2i\gamma^7D^I_t)e^{\sigma}\sigma^t_{AB}-
2me^{-3\sigma}(L^{-1})^I_{\ \ 0}\epsilon_{AB}, \label{mgM}
\end{eqnarray}
 besides the new definition $\hat F
\equiv F -mB$.\\ We thus obtain the supersymmetry transformation
rules in presence of gauging and with the mass parameter turned
on:
\begin{eqnarray}
&\delta
V^{a}_{\mu}&=-i\overline{\psi}_{A\mu}\gamma^{a}\varepsilon^A\nonumber\\
&\delta B_{\mu\nu}&=4\ii e^{-2\sigma}
\overline{\chi}_{A}\gamma_{7}\gamma_{\mu\nu}\varepsilon^A
-4e^{-2\sigma}\overline{\varepsilon}_A\gamma_7\gamma_{[\mu}\psi_{\nu]}^A\nonumber\\
&\delta A^{\Lambda}_{\mu}&=2 e^{\sigma}
 \overline{\varepsilon}^{A}\gamma_{7}\gamma_{\mu}\chi^BL^{\Lambda}_0
 \epsilon_{AB}+2e^{\sigma}\overline{\varepsilon}^{A}\gamma_{\mu}\chi^{B}
 L^{\Lambda r}\sigma_{rAB}-e^{\sigma}L_{\Lambda
I}\overline{\varepsilon}^{A}\gamma_{\mu}\lambda^{IB}\epsilon_{AB}+\nonumber\\
&&+2ie^{\sigma}L^{\Lambda}_0\overline{\varepsilon}_A\gamma^7\psi_B\epsilon^{AB}+
2ie^{\sigma}L^{\Lambda r}\sigma_{r}^{AB}\overline{\varepsilon}_A\psi_B\nonumber\\
&\delta\psi_{A\mu}&=\mathcal{D}_{\mu}\varepsilon_A+\frac{1}{16}
e^{-\sigma}[\hat{T}_{[AB]\nu\lambda}\gamma_{7}-{T}_{(AB)\nu\lambda}](\gamma_{\mu}^{\,\
\nu\lambda}-6\delta_{\mu}^{\nu}\gamma^{\lambda})
\varepsilon^{B}+\nonumber \\
&&+\frac{i}{32}e^{2\sigma} H_{\nu\lambda\rho}
\gamma_{7}(\gamma_{\mu}^{\,\ \nu\lambda\rho}-3\delta_{\mu}^{\nu}
\gamma^{\lambda\rho})\varepsilon_{A}+ S^{(g,g',m)}_{AB}
\gamma_\mu\varepsilon^B +\nonumber\\
&&+\frac{1}{2}\varepsilon_{A}\overline{\chi}^{C}\psi_{C\mu}+
\frac{1}{2}\gamma_{7}\varepsilon_{A}\overline{\chi}^{C}\gamma^{7}\psi_{C\mu}-
\gamma_{\nu}\varepsilon_{A}\overline{\chi}^{C}\gamma^{\nu}\psi_{C\mu}+\gamma_{7}
\gamma_{\nu}\varepsilon_{A}\overline{\chi}^{C}\gamma^{7}\gamma^{\nu}\psi_{C\mu}+\nonumber\\
&&-\frac{1}{4}\gamma_{\nu\lambda}\varepsilon_{A}\overline{\chi}^{C}\gamma^{\nu\lambda}\psi_{C\mu}-
\frac{1}{4}\gamma_{7}\gamma_{\nu\lambda}\varepsilon_{A}\overline{\chi}^{C}\gamma^{7}\gamma^{\nu\lambda}\psi_{C\mu}
\nonumber\\
&\delta\chi_{A}&=\frac{i}{2} \gamma^{\mu}\partial_{\mu}\sigma
\varepsilon_{A}+
\frac{i}{16}e^{-\sigma}[\hat{T}_{[AB]\mu\nu}\gamma_{7}+T_{(AB)\mu\nu}]\gamma^{\mu\nu}\varepsilon^{B}+\frac{1}{32}e^{2\sigma}
H_{\mu\nu\lambda}\gamma_{7}\gamma^{\mu\nu\lambda}\varepsilon_{A} + \nonumber\\
&&+ N^{(g,g',m)}_{AB}\epsilon^B \nonumber\\
&\delta\sigma&=\overline{\chi}_{A}\varepsilon^A\nonumber\\
&\delta\lambda^{I}_A&=i\hat{P}^I_{ri}\sigma^{r}_{AB}\partial_{\mu}\phi^{i}\gamma^{\mu}\varepsilon^{B}-
i\hat{P}^I_{0i}\epsilon_{AB}\partial_{\mu}\phi^{i}\gamma^{7}\gamma^{\mu}\varepsilon^{B}+
\frac{i}{2}e^{-\sigma}T^{I}_{\mu\nu}\gamma^{\mu\nu}\varepsilon_{A} + M^{I (g,g',m)}_{AB}\varepsilon^B \nonumber \\
&\hat{P}^{I}_{0i}\delta\phi^i&=\frac{1}{2}\overline{\lambda}^{I}_{A}\gamma_{7}\varepsilon^A\nonumber\\
&\hat{P}^{I}_{ri}\delta\phi^i&=\frac{1}{2}\overline{\lambda}^{I}_{A}\varepsilon_{B}\sigma_r^{AB}
\label{newsusy}
\end{eqnarray}
where we have introduced the "dressed" non abelian vector field
strengths:
\begin{eqnarray}
&&\hat{T}_{[AB]\mu\nu}\equiv\epsilon_{AB}L^{-1}_{0\Lambda
}\left(F^{\Lambda}_{\mu\nu} -m B_{\mu\nu} \delta_{\Lambda 0}\right)\\
&&T_{(AB)\mu\nu}\equiv\sigma^r_{AB}L^{-1}_{r\Lambda}F^{\Lambda}_{\mu\nu}\\
&&T_{I\mu\nu}\equiv L^{-1}_{I\Lambda }F^{\Lambda}_{\mu\nu}
\end{eqnarray}
the gauged scalar vielbein $\widehat{P}^I_0$, $\widehat{P}^I_r$
being defined in eq. \eq{scalviel}.

From the transformation laws  \eq{newsusy}, it is easy to see that
one can obtain an Anti de Sitter supersymmetric background
choosing $g=3m$. (Recall that in the vacuum, besides putting to
zero the field-strengths, we also set $\s =0$, $L^\L_{~\Sigma} =
\delta^\L_\Sigma$). In this way we obtain:
\begin{eqnarray}
&&\delta\chi_A\equiv -\frac{1}{4} \left(g-3m\right)\varepsilon_A= 0\\
&&\delta\psi_{A\mu}\equiv D_\mu \varepsilon_A -\ii (\frac{3}{4}m+\frac{1}{4}g)\gamma_\mu \varepsilon_A=D_\mu \varepsilon_A -\ii m \gamma_\mu \varepsilon_A=\nabla^{AdS}_{\mu}\epsilon_A\\
\label{curv}&&R^{ab}\equiv -\frac{1}{2} R^{ab}_{cd}V^cV^d=-4m^2 V^aV^b\rightarrow R_{\mu\nu}=20m^2g_{\mu\nu}\\
&&\left(\mathcal{F}^r_{\mu\nu}=\mathcal{F}_{\mu\nu}-mB_{\mu\nu}=H_{\mu\nu\rho}=\chi_A=\psi_{A\mu}=\sigma=0;\,L^\L_{~\Sigma}
= \delta^\L_\Sigma \right)
\end{eqnarray}
 which corresponds to an $AdS$ configuration with $AdS$ radius
$R^2_{AdS}=(4m^2)^{-1}$
%%%%%%%%%%%%%%%%%%%%%%%%%%%%%%%%%%%%%%%%%%%%%%%%%%%%%%%%%%%%%%%%%%%%%%%%%%%%%%%%%%%%%%%%%%%%%%
%%%%%%%%%%%%%%%%%%%%%%%%%%%%%%%%%%%%%%%%%%%%%%%%%%%%%%%%%%%%%%%%%%%%%%%%%%%%%%%%%%%%%%%%%%%%%%

\section{The complete gauged lagrangian}
\setcounter{equation}{0}
 When the gauging is present the
Lagrangian can be constructed by covariantization (with respect to
the gauge group) of all the derivatives present in the ungauged
one \eq{lagrungauged} plus extra terms, bilinears in the fermions,
proportional to the coupling constants $g$, $g'$ and $m$ and a
scalar potential which is quadratic in the same parameters. As
for the ungauged case, the explicit construction has been done
using the rheonomic formalism in superspace and taking advantage
of the paramentrizations of the curvatures obtained by solving
the Bianchi identities. A short account of this is given in
Appendix B. By restricting the superspace Lagrangian to
space--time we get: \be {\mathcal A}_{(D=6,N=2)} = \int
\mathcal{L}^{(g,g',m)}_{(D=6,N=2)} \sqrt{-g} \, d^6 x ,  \ee
 \be
  \mathcal{L}^{(g,g',m)}_{(D=6,N=2)}\equiv
\mathcal{L}_{\mbox{\tiny{kin}}}
+\mathcal{L}_{\mbox{\tiny{Pauli}}}+
\mathcal{L}_{\mbox{\tiny{Chern--Simons}}}+
\mathcal{L}_{\mbox{\tiny{gauging}}} +\mathcal{L}_{\mbox{\tiny{4
fermions}}} \label{lagrgauged} \ee where:
%%%%%%%%%%%%%%%%%%%%%%%%%%%%%%%%%%%%%%%%%%%%%%%%%%%%%
%%%%%%%%%lagrangiana cinetica %%%%%%%%%%%%%%%%%%%%%
%%%%%%%%%%%%%%%%%%%%%%%%%%%%%%%%%%%%%%%%%%%%%%%%%%%%
 \ba \mathcal{L}_{\mbox{\tiny{kin}}}&=&
-\frac{1}{4}\mathcal{R}-\frac{1}{8}e^{-2\sigma}\mathcal{N}_{\Lambda\Sigma}
\widehat{\mathcal{F}}^{\Lambda}_{\mu\nu}\widehat{\mathcal{F}}^{\Sigma\mu\nu}
+\frac{3}{64}e^{4\s}H_{\mu\nu\rho}H^{\mu\nu\rho}+\nonumber\\
&&+\frac{i}{2}\overline{\psi}_{A\mu}\gamma^{\mu\nu\rho}
\nabla_{\nu}\psi^{A}_{\rho} -
2i\overline{\chi}_A\gamma^{\mu}\nabla_{\mu}\chi^A-
\frac{i}{8}\overline{\lambda}^I_A\gamma^{\mu}\nabla_{\mu}\lambda^A_I+\nonumber\\
&&+
\partial^{\mu}\sigma\partial_{\mu}\sigma  -\frac{1}{4}
\left(\widehat{P}^{I0}_{ i} \widehat{P}_{I0 j}
+\widehat{P}^{Ir}_{i}\widehat{P}_{Ir , j}\right)
\partial^{\mu}\phi^i\partial_{\mu}\phi^{j}
 \, ;\ea
%%%%%%%%%%%%%%%%%%%%%%%%%%%%%%%%%%%%%%%%%%%%%%%%%%%%%
%%%%%%%%%lagrangiana di pauli %%%%%%%%%%%%%%%%%%%%%
%%%%%%%%%%%%%%%%%%%%%%%%%%%%%%%%%%%%%%%%%%%%%%%%%%%%
\ba {\mathcal{L}}_{\mbox{\tiny{Pauli}}}&=& -2\partial_{\mu}\s
\bar\chi_A\g^{\nu}\g^{\mu}\p^A_{\nu}
+\frac{1}{4}\widehat{P}^{I0}_{i}\partial^{\mu}\phi^i
\bar\lambda_{IA}\g_7\g^{\nu}\g^{\mu}\p^A_{\nu}
+\frac{1}{4}\widehat{P}^{I(AB)}_{i}\partial^{\mu}\phi^i
\bar\lambda_{IA}\g^{\nu}\g^{\mu}\p_{B\nu} + \nonumber\\
&&+e^{-\s}\mathcal{N}_{\L\Sigma}\hat{\mathcal{F}}^{\L}_{\mu\nu}
\Big\{ \frac{i}{8}L^{\Sigma}_0\pb_A^{\rho}
\g_7\left(\g^{\mu\nu}_{\phantom{\mu\nu}\rho\s}+2\delta^{\mu\nu}_{\rho\s}\right)\p^{A\s}+
 \frac{i}{8}L^{\Sigma}_{(AB)}\pb^{A\rho}
 \left(\g^{\mu\nu}_{\phantom{\mu\nu}\rho\s}+2\delta^{\mu\nu}_{\rho\s}\right)\p^{B\s}+\nonumber\\
&&-\frac{1}{4}L^{\Sigma}_0
\pb_{A\rho}\g_7\g^{\mu\nu}\g^{\rho}\chi^A
-\frac{1}{4}L^{\Sigma}_{(AB)}\pb_{\rho}^A\g^{\mu\nu}\g^{\rho}\chi^B
-\frac{1}{8}L^{\Sigma}_I
\bar\lambda^I_A\g^{\mu\nu}\g^{\rho}\psi^A_{\rho}+\nonumber\\
&&+L^{\Sigma}_0
\left(\frac{\ii}{4}\bar\chi_{A}\g_7\g^{\mu\nu}\chi^A+\frac{\ii}{32}\bar\lambda_{IA}\g_7\g^{\mu\nu}\l^{IA}\right)
+L^{\Sigma}_{(AB)}
\left(-\frac{\ii}{4}\bar\chi^{A}\g^{\mu\nu}\chi^B+\frac{\ii}{32}\bar\lambda_{I}^{A}\g^{\mu\nu}\l^{IB}\right)
\Big\}+\nonumber\\
 &&+\frac{3}{8}e^{2\s}H_{\mu\nu\rho}
\Big\{\frac12\pb_A^{\l}\g_7
\left(\delta^{\mu\nu}_{\l\s}\g^{\rho}-16\g^{\mu\nu\rho}_{\phantom{\mu\nu\rho}\l\s}\right)\p^{A\s}
+\frac{i}{3}\pb_{A\s}\g_7\g^{\s}\g^{\mu\nu\rho}\chi^A +\nonumber\\
&&+\frac{i1}{3}\bar\chi_A\g_7\g^{\mu\nu\rho}\chi^A\Big\}\, ;\ea
%%%%%%%%%%%%%%%%%%%%%%%%%%%%%%%%%%%%%%%%%%%%%%%%%%%%%
%%%%%%%%%lagrangiana di chern-simons %%%%%%%%%%%%%%%
%%%%%%%%%%%%%%%%%%%%%%%%%%%%%%%%%%%%%%%%%%%%%%%%%%%%
\ba
 \mathcal{L}_{\mbox{\tiny{Chern--Simons}}}&=&
 -\frac{1}{64}\epsilon^{\mu\nu\rho\s\l\tau}B_{\mu\nu}
\left(\eta_{\L\Sigma}\hat{\mathcal{F}}^{\L}_{\rho\s}\hat{\mathcal{F}}^{\Sigma}_{\l\tau}
+mB_{\rho\s}\hat{\mathcal{F}}^0_{\l\tau}+\frac{1}{3}m^2B_{\rho\s}B_{\l\tau}\right)
\ea
%%%%%%%%%%%%%%%%%%%%%%%%%%%%%%%%%%%%%%%%%%%%%%%%%%%%%
%%%%%%%%%lagrangiana di gauging %%%%%%%%%%%%%%%%%%%
%%%%%%%%%%%%%%%%%%%%%%%%%%%%%%%%%%%%%%%%%%%%%%%%%%%%
\ba
 \mathcal{L}_{\mbox{\tiny{gauging}}}&=&
2\ii\overline{\psi}_{\mu}^A\gamma^{\mu\nu}\overline{S}_{AB}\psi_{\nu}^B+
+4\ii\overline{\psi}_{\mu}^A\gamma^{\mu}\overline{N}_{AB}\chi^{B}-\frac{\ii}{4}
\overline{\psi}_{\mu}^A\gamma^{\mu}\overline{M}_{AB}^I\lambda^{B}_I+\nonumber\\
&&+ \bar\chi^A X_{AB} \chi^B + \bar\chi^A Y^I_{AB} \lambda_I^B +
\bar\lambda^I_A Z^{AB}_{IJ}\lambda^J_B -
 \mathcal{W}(\sigma , \, \phi^i;g,g',m) .\label{lgaug}\ea
where the covariant derivatives in
$\mathcal{L}_{\mbox{\tiny{kin}}}$ are now Lorentz- and
$SU(2)\otimes \cG$-covariant derivatives. ${\cal L}_{\mbox{\tiny{4
fermions}}}$ has not been computed.
\par
In equation (\ref{lgaug}) there appear ``barred mass-matrices''
$\overline{S}_{AB},\,\ \overline{N}_{AB},\,\ \overline{M}^I_{AB}$
which are slightly different from the fermionic shifts defined in
eqs. (\ref{mS}), (\ref{mN}), (\ref{mM}). Actually they are
defined by:
\begin{equation}
\label{pippo}\overline{S}_{AB}=-S_{BA},\ \ \ \
\overline{N}_{AB}=-N_{BA},\ \ \ \ \overline{M}^I_{AB}=M^I_{BA}
\end{equation}
 Definitions (\ref{pippo}) stem from the fact that the shifts
defined in eqs. (\ref{mS}), (\ref{mN}), (\ref{mM}) are matrices
in the eight-dimensional spinor space, since they contain the
$\gamma_7$ matrix; as will be seen in a moment, such definitions
are actually necessary in order to satisfy the
supersymmetry Ward identities.\\
Furthermore, the mass matrices of the spin one-half fermions and
the potential can be computed from supersymmetry of the
Lagrangian and turn out to be:
\begin{eqnarray}
 X_{AB}&=& 4{\ii} \left(S_{AB} -2\ii \overline{N}_{AB}
 \right)\nonumber\\
 Y^I_{AB}&=& -\frac{1}{2}\partial_{\s}\overline{M}^I_{AB}\nonumber\\
Z^{AB}_{IJ} &=& \frac{\ii}{4}\left[\left(\overline{S}^{AB}+\ii
N^{AB}\right)\eta_{IJ}+
\left(K_{rIJ}\s^{rAB}-K_{0IJ}\g_7\epsilon^{AB}\right)\right]
\label{massespinori}
\\
\mathcal{W}(\phi)&=& - 5\,\ \{
[\frac{1}{12}(Ae^{\sigma}+6me^{-3\sigma}L_{00})]^2+[\frac{1}{4}(e^{\sigma}B_r-2me^{-3\sigma}L_{0r})]^2\}+\nonumber\\
&&+ \{
[\frac{1}{12}(Ae^{\sigma}-18me^{-3\sigma}L_{00})]^2+[\frac{1}{4}(e^{\sigma}B_r+6me^{-3\sigma}L_{0r})]^2\}+\nonumber\\
&&-\frac{1}{4}\{C^I_{\,\ t}C_{It}+4D^I_{\,\ t}D_{It}\}\,\
e^{2\sigma} - m^2e^{-6\sigma}L_{0I}L^{0I} \label{pot}
\end{eqnarray}
It is convenient to discuss in detail the determination of the
potential from the supersymmetry variation of the lagrangian.
Indeed, let us perform the supersymmetry variation of
(\ref{lagrgauged}), keeping only the terms quadratic in $g$, $g'$
or $m$, proportional to the currents
$\overline{\psi}_{A\mu}\gamma^{\mu}\epsilon^C$ and
$\overline{\psi}_{A\mu}\g_7\gamma^{\mu}\epsilon^C$; we find the
following Ward identity, :
\begin{equation}
\label{ward} -\delta^A_C \, \mathcal{W} \,
\overline{\psi}_{A\mu}\gamma^{\mu}\epsilon^C=
\overline{\psi}_{A\mu}\gamma^{\mu} \left(
20\overline{S}^{AB}S_{BC}+4\overline{N}^{AB}N_{BC}+\frac{1}{4}\overline{M}^{AB}_IM^I_{BC}\right)\epsilon^C
\end{equation}
Note that, on the right-hand-side, the terms quadratic in the
shifts give  rise to terms proportional to the current
$\overline{\psi}_{A\mu}\gamma^7\gamma^{\mu}\epsilon^A$, which
have no counterpart in the term containing the potential $\cW$
and therefore must cancel against each other. This is actually
what happens for the first two terms on the r.h.s. of (\ref{ward})
taking into account the definition of the barred mass matrices
given in eq. (\ref{pippo}).
 As far as the term
$\overline{M}^{AB}_IM^I_{BC}$ is concerned,
 the same mechanism of cancellation again applies to the terms
proportional to
$\overline{\psi}_{A\mu}\gamma_7\gamma^{\mu}\epsilon^A\sigma^{rA}_C$;
there is, however, a residual dangerous term of the form
\begin{equation}
\delta^A_C\overline{\psi}_{A\mu}\gamma^{\mu}\gamma^7D^I_{\,\
s}C_I^{\,\ s}\epsilon^C
\end{equation}
\noindent One can show that this term vanishes identically owing
to the non trivial relation
\begin{equation}
\label{abigaille}D^I_tC_I^t=0
\end{equation}
Equation (\ref{abigaille}) can be shown to hold  using the
pseudo-orthogonality relation $L^T\eta L=\eta$ among the coset
representatives and the Jacobi identities $C_{I[JK}C_{L]MN}=0$,
$\epsilon_{r[st}\epsilon_{l]mn}=0$. This is a non trivial check of our computation.\\
Note that, setting
\begin{eqnarray}
&&\mathcal{H}=\frac{1}{12}(Ae^{\sigma}+6me^{-3\sigma}L_{00})\\
&&\mathcal{K}_i=\frac{1}{4}(e^{\sigma}B_i-2me^{-3\sigma}L_{0i})
\end{eqnarray}
the potential can be written as follows
\begin{eqnarray}
\label{bibo}&&\mathcal{W}=-5\{\mathcal{H}^2+\mathcal{K}^i\mathcal{K}_i\}+
\{\partial_{\sigma}\mathcal{H}\partial_{\sigma}\mathcal{H}+\partial_{\sigma}\mathcal{K}^i\partial_{\sigma}\mathcal{K}_i\}
+2\{\nabla_{I\alpha}\mathcal{H}\nabla^{I\alpha}\mathcal{H}+\nabla_{I\alpha}\mathcal{K}_i\nabla^{I\alpha}\mathcal{K}^i\}+\nonumber\\
&&+m^2e^{-6\sigma}L_{0I}L^{0I}
\end{eqnarray}
\noindent where $\nabla_{I\alpha}\equiv
(\nabla_{I0},\nabla_{Ir})$ denote the derivatives with respect to
the "linearized coordinates": that is, using the Maurer-Cartan
equations
\begin{eqnarray}
&&\nabla^H L^{\Lambda}_{\ \ I}=L^{\Lambda}_{\ \
\alpha}P^{\alpha}_I\nonumber\\ \label{ugo}&&\nabla^H
L^{\Lambda}_{\ \ \alpha}=L^{\Lambda}_{\ \ I}P_{\alpha}^I
\end{eqnarray}
 the flat derivative $\nabla_{I\alpha}$ are defined as
the coefficient of the coset vielbein $P^{I\alpha}$ in equations
(\ref{ugo}). In deriving equations
(\ref{massespinori})--(\ref{bibo}) one has to make use of the
following relations which are a straightforward consequence of
the definitions (\ref{AA}) -(\ref{DD})
\begin{eqnarray}
&&\nabla_{I0}A=0\\
&&\nabla_{Ir}A=3C_{Ir}\\
&&\nabla_{I0}B_i=C_{iI}\\
&&\nabla_{Ir}B_i=2\epsilon_{rik}D_{Ik}.
\end{eqnarray}
Expanding the squares in equation (\ref{pot}) the potential can be
alternatively written as follows:
\begin{eqnarray}
&&\mathcal{W}=-e^{2\sigma}[\frac{1}{36}A^2+\frac{1}{4}B^iB_i+\frac{1}{4}(C^I_{\,\
t}C_{It}+4D^I_{\,\
t}D_{It})]+m^2e^{-6\sigma}\mathcal{N}_{00}+\nonumber\\
&&-me^{-2\sigma}[\frac{2}{3}AL_{00}-2B^iL_{0i}]
\end{eqnarray}
where $\mathcal{N}_{00}$ is the 00 component of the vector
kinetic matrix defined as in equation \eq{kin}.

We now show that, apart from other possible extrema not
considered here, a stable supersymmetric extremum of the potential
$\mathcal{W}$ is found to be the same as in the case of pure
supergravity, that is we get an $AdS$ supersymmetric background
only for $g=3m$. In fact, setting
$\partial_{\sigma}\mathcal{W}=0$ and keeping only the non
vanishing terms at $\sigma=q^{I\alpha}=0$, $q^{I\alpha}$ being
the flat coordinates (that is the coordinates associated to the
flat derivatives, or equivalently the coordinates of the
linearized theory) we have
\begin{eqnarray}
&&\partial_{\sigma}\mathcal{W}=[\frac{1}{18}A^2e^{2\sigma}-
\frac{4}{3}mAL_{00}e^{-2\sigma}+6m^2L_{00}^2e^{-6\sigma}]_{\sigma=q^{I\alpha}=0}
\end{eqnarray}
\noindent since all the other terms entering the
$\partial_{\sigma}\cW$ contain at least one off-diagonal element
of the coset representative which vanishes identically when the
scalar fields are set equal to zero. Furthermore, from the
definition (\ref{AA}) and using $L^{\Lambda}_{\,\
\Sigma}(q^I_{\alpha}=0)=\delta^{\Lambda}_{\Sigma}$, we find:
\begin{equation}\label{aq}A(q^I_{\alpha}=0)=6g;\ \ \ \ L_{00}(q^I_{\alpha}=0)=1\end{equation}
\noindent so that
\begin{eqnarray}
\label{pluto}&&\partial_{\sigma}\mathcal{W}|_{\sigma=q=0}=2g^2-8mg+6m^2=0
\end{eqnarray}
As the partial derivatives $(\frac{\partial\mathcal{W}}{\partial
q^{I0}})_{\sigma=q=0}$, $(\frac{\partial\mathcal{W}}{\partial
q^{Ir}})_{\sigma=q=0}$ are also zero, since they contains at
least one off-diagonal coset representative, the condition for
the minimum is given by eq. (\ref{pluto}) which coincides with
the equation one obtains for
the pure Supergravity case, whose solutions are $g=m$, $g=3m$.\\
Using equations (\ref{aq}), (\ref{pluto}), (\ref{del1})
-(\ref{del3}) one  can  easily recognize  that only
the $g=3m$ solution gives rise to a supersymmetric $AdS$ background.\\
An important point regarding this Lagrangian is that it
incorporates the Higgs mechanism for the two-form $B$ as it was
first noticed in the pure supergravity case by Romans \cite{rom}.
Indeed, the field-strength for the $SU(2)$--singlet vector
$A^0_\mu$ only appears in the Lagrangian in the combination
$F^0_{\mu\nu} -mB_{\mu\nu}$. By performing the gauge
transformation \ba \delta B_{\mu\nu} &=& \partial_{[\mu} f_{\nu]}
\nonumber\\ \delta A^0_\mu &=&  mf_\mu \ea which leaves
$\widehat{\cF}^0_{\mu\nu}$ invariant, we can choose $f$ in such a
way that the field $A^0_\mu$ disappears from the theory, leaving
$\widehat{\cF}^0_{\mu\nu}=-mB_{\mu\nu}$ so that the Lagrangian
could be rewritten without any reference to the singlet-vector
field $A^0_\mu$, but with a massive two--form $B_{\mu\nu}$. Of
course this is in complete agreement with what we found in section
2 at the level of the F.D.A. on which the theory is based.
\par
Incidentally, we note that besides the Ward identity \eq{ward}
there are further supersymmetry Ward identities which relate the
gradient of the fermionic shifts with themselves and the spin
one-half mass--matrices. They are pretty analogous to the
gradient flow equations studied in ref. \cite{df} for $N=1$ and
$N=2$ $D=4$ supergravity. The $D=6$ gradient flows are the
following:
%%%%%%%%%%%%%%%%%%%%%%%%%%%%%%%%%%%%%%%%%%%%%%%%%%%%%%%%%%%%%%%%%%%%%%%%%%%%%%%%%%%%%%%%%%%%%%
%%%%%%%%% relazioni tra le matrici di massa %%%%%%%%%%%%%%%%%%%%%%%%%%%%%%%%%%%%%%%%%%%%%%%%%%
%%%%%%%%%%%%%%%%%%%%%%%%%%%%%%%%%%%%%%%%%%%%%%%%%%%%%%%%%%%%%%%%%%%%%%%%%%%%%%%%%%%%%%%%%%%%%%
 \ba
\partial_{\s}S_{AB}&=&\ii\overline{N}_{AB}\\
\nabla_{I0}S_{AB} &=& -\frac{\ii}{4}m e^{-3\s}L_{0I}\epsilon_{AB}
-\frac{\ii}{8} C_{Ir}e^\s \g_7 \s^r_{AB}\\
\nabla_{Is}S_{AB} &=& -\frac{\ii}{4}m e^{-3\s}L_{0I}\g_7 \s_{s
|AB}+\frac{\ii}{8} C_{Is}e^\s \epsilon_{AB}+\frac{\ii}{4}
\epsilon_{rst}D_I^r e^\s \g_7 \s^t_{AB}
\\
\partial_{\s}\overline{N}_{AB}&=&-\frac{1}{4}X_{AB}-2\ii S_{AB}=-2\bar{N}_{AB}-3\ii S_{AB}\\
\nabla_{I0}N_{AB} &=& -\frac{3}{4}m e^{-3\s}L_{0I}\epsilon_{AB}
+\frac{1}{8} C_{Ir}e^\s \g_7 \s^r_{AB}\\
\nabla_{Is}N_{AB} &=& -\frac{3}{4}m e^{-3\s}L_{0I}\g_7 \s_{s
|AB}+\frac{1}{8} C_{Is}e^\s  \epsilon_{AB}-\frac{1}{4}
\epsilon_{rst}D_I^r e^\s \g_7 \s^t_{AB}
\\
\partial_{\s}\bar{M}^I_{AB}&=&-2Y^I_{AB}=8\ii
(\sigma^{sC}_B\nabla_{Is}S_{AC}-\gamma_7\nabla_{I0}S_{AB})-M_{IAB}
\ea The previous formulae can be important in the study of
renormalization group flows related to the scalar potential of
our theory.
\par
A further issue, which is an important check of all our
calculation, is the possibility of computing the masses of the
gravitational and vector supermultiplets in the Anti de Sitter
background. First we compute the masses of the scalar fields by
varying the linearized kinetic terms and the potential of
(\ref{lagrgauged}), after power expansion of $\mathcal{W}$ up to
the second order in the
scalar fields $q^I_{\alpha}$.\\
We find:
\begin{eqnarray}
&&(\frac{\partial^2\mathcal{W}}{\partial\sigma^2})_{\sigma=q=0,
g=3m}=48m^2\\
&&(\frac{\partial^2\mathcal{W}}{\partial q^{I0}\partial q^{J0}}
)_{\sigma=q=0,
g=3m}=8m^2\delta^{IJ}\\
&&(\frac{\partial^2\mathcal{W}}{\partial q^{Ir}\partial q^{Js}
})_{\sigma=q=0, g=3m}=12 m^2\delta^{IJ}\delta^{rs}
\end{eqnarray}
The linearized equations of motion become:
\begin{eqnarray}
&&\Box\sigma-24m^2\sigma + \cdots =0\\
&&\Box q^{I0}-16m^2 q^{I0} + \cdots =0\\
&&\Box q^{Ir}-24m^2 q^{Ir} + \cdots =0
\end{eqnarray}
If we use as mass unity the
\def\IP{\relax{\rm I\kern-.18em P}} inverse $AdS$ radius, which
in our conventions (see eq.(\ref{curv})) is $R^{-2}_{AdS}=4m^2$
we get:
\begin{eqnarray}
&&m^2_{\sigma}=-6\nonumber\\
&&m^2_{q^{I0}}=-4\nonumber\\
\label{massa}&&m^2_{q^{Ir}}=-6
\end{eqnarray}
In the same way we may compute the masses of all the other fields
of the multiplets. One finds the following  linearized equations
of motion: \ba
 \ii \g^\mu \nabla_\mu \l^A_I + 4 m^2 \l^A_I + \cdots &=& 0\nonumber\\
 \ii \g^\mu \nabla_\mu \chi^A +4 m^2 \chi^A + \cdots &=& 0\nonumber\\
  \ii \g^\mu \nabla_\mu \psi^A + 8 m^2 \psi^A + \cdots &=& 0 \nonumber\\
\nabla_\mu \widehat{\cF}^{\mu\nu} + \cdots &=& 0 \nonumber\\
 \nabla_\mu \cH^{\mu\nu\rho} +\frac 83 m^2 B^{\nu\rho}+ \cdots &=& 0\nonumber\\
 \ea
 so that, in units of $R^{-2}_{AdS}$ we get:
 \begin{eqnarray}
 && m_\p = 2\, , \,
 m_\l =
m_\chi = 1\nonumber\\
&& m_{A^I} = m_{A^\alpha} = 0 \, , \, m^2_{B_{\mu\nu}} = 2
\label{masseferm}
\end{eqnarray}
 These values should be compared with the results
obtained in reference \cite{fkpz} where the supergravity and
matter multiplets of the $AdS_6\,\ F(4)$ theory were constructed
in terms of the singleton fields of the 5-dimensional conformal
field theory, the singletons being given by hypermultiplets
transforming in the fundamental of $\mathcal{G}\equiv E_7$. It is
amusing to see that the values of the masses of all the fields
computed in terms of the conformal dimensions are exactly the
same as those given in equation (\ref{massa}), \eq{masseferm}.
Indeed, using the relations between $E_0$ and the masses as given
for example in \cite{fgpw}:
\begin{eqnarray}
m^2 =& E_0 \left(E_0 -5\right)
 \,\, &\mbox{for scalars}\nonumber\\
|m| =& E_0 - \frac 52
\, \, &\mbox{for $\half$-spinors}\nonumber\\
m^2 =& \left(E_0 -1\right)\left(E_0-4\right)
\,\, &\mbox{for vectors}\nonumber\\
m^2 =& \left(E_0 -2\right)\left(E_0-3\right)
\,\, &\mbox{for 2-forms}\nonumber\\
|m| =& E_0 - \frac 52
 \,\, &\mbox{for $\frac 32$-spinors}\nonumber\\
m^2 =& E_0 \left(E_0 -5\right)  \,\, &\mbox{for graviton}
\end{eqnarray}
it is immediate to retrieve for the masses of all the fields of
the supermultiplets, using Table 1, the values appearing in the
above linearized equations of motion.
 This coincidence can be considered as a non
trivial check of the
$AdS/CFT$ correspondence in six versus five dimensions.\\
\begin{table}[h]
\begin{tabular}{l|ccccccc}
\hline \hline
&&&&&&&\\
 Grav. mult. & $ g_{\mu\nu}$ & $\psi_A$ & $A^0_\mu$ & $A^r_\mu$ &
 $B_{\mu\nu}$ & $ \chi_A$ & $e^\sigma$ \\
\hline &&&&&&&\\
 $E_0$ & 5&$\frac 92 $ & 4&4&4& $\frac 72 $ & 3 \\
%&&&&&&&\\
 \hline
 \hline
&&&&&&&\\
 Vector mult. &$A^I_\mu$ & $\l^A_I $& $q^{I0}$& $q^{Ir}$ & && \\
\hline
&&&&&&&\\
$E_0$ & 4&$\frac 72 $ & 4&3&&&  \\
%&&&&&&&\\
\hline
\end{tabular}
\caption{$E_0$ values for the gravitational and vector
multiplets, respectively.}
\end{table}

Let us finally observe that the scalar squared masses in
$AdS_{d+1}$ are given by the $SO(2,d)$ quadratic Casimir
\cite{flafro} \be m^2=E_0(E_0-d). \ee They are negative in the
interval $\frac{d-2}{2}\leq E_0<d$ (the lower bound corresponding
to the unitarity bound i.e. the singleton) and attain the
Breitenlohner-Freedman bound \cite{breit} when $E_0=d-E_0$ i.e.
at $E_0=\frac{d}{2}$ for which $m^2=-\frac{d^2}{4}$. Conformal
propagation corresponds to $m^2=-\frac{d^2-1}{4}$ i.e.
$E_0=\frac{d\pm 1}{2}$. This is the case of the dilaton and
triplet matter scalars.

%%%%%%%%%%%%%%%%%%%%%%%%%%%%%%%%%%%%%%%%%%%%%%%%%%%%%%%%%%%%%%%%%%%%
%%%%%%%%%%%%%%%%%%%%%%%%%%%%%%%%%%%%%%%%%%%%%%%%%%%%%%%%%%%%%%%%%%%%
\section{Conclusions}
In this paper we have given the so far unknown complete
Lagrangian (up to four-fermion terms) of the matter coupled $D=6$
$F(4)$ supergravity.
\par
Besides to fill a gap in the supergravity literature, this is a
necessary step in order to perform a complete analysis of the
$AdS_6/CFT_5$ correspondence. Indeed, in ref. \cite{noi} such
correspondence has been established only as far as the
supersymmetric general structure of the vector multiplets and
gravitational multiplet is concerned. In particular, it was found
that there are two series of unitary irreducible representations
in the five dimensional superconformal field theory which
correspond to a massive tower of short vector multiplets and to a
tower of massive graviton multiplets respectively. The lowest
members of the two towers are actually massless, and correspond
to the conserved currents of the global flavour ${\cal G}$
symmetry of the five dimensional conformal field theory and to the
stress--tensor multiplet respectively. However, in that framework
it was not possible to determine the ${\cal G}$ quantum numbers
of the supermultiplets. On the other hand, in ref. \cite{oz} it
was established that the $F(4)$ supergravity theory can be
obtained as a Kaluza--Klein reduction of massive Type IIA
supergravity on a background $AdS_6$ which is fibered with a
warped four--sphere. This reduction is related to the horizon
geometry of $D4$-branes in a $D8$-brane background, in presence
of a $D0$-brane. In the same reference the warped metric for the
reduction was obtained. In particular, in a recent paper
\cite{canu}, solutions for Romans' six-dimensional gauged
supergravity, related to D-branes interpretations, were found and
then lifted up to ten dimensions.
\\
The knowledge of the $F(4)$ Lagrangian in $D=6$ together with
that of the warped metric for the compactifiction, allows one to
perform in principle an analysis of the complete Kaluza--Klein
spectrum, which would exhibit, besides the supermultiplet
structure, also the ${\cal G}$ flavour quantum numbers. This
program is left to a future investigation.
\section*{Acknowledgements}
We thank Sergio Ferrara for useful discussions during the
completion of the paper. This work has been supported by the
European Commission RTN network HPRN-CT-2000-00131 (Politecnico
di Torino).
\appendix
\section*{Appendix A: The superspace curvature and the solution of Bianchi identities.}
\setcounter{equation}{0} \label{appendiceA}
\addtocounter{section}{1} As it has been stressed in the text,
the supersymmetry transformation laws for the physical fields
have been obtained by solving the Bianchi identities in
superspace. In our case we follow the particular approach
developed in \cite{bible}, which is substantially equivalent to
the other existing superspace approaches, the only difference
being the more precise group--theoretical assessment of the
starting points.
\par
The first step is to find the solution of Bianchi identities
\eq{xtors} -- \eq{xdP} in absence of gauging.
\par
The solution can be obtained as follows: first of all one
requires that the expansion of the curvatures along an intrinsic
$p$--form basis in superspace generated by $V^a$ and $\psi$  is
given in terms only of the physical fields. This insures that no
new degree of freedom is introduced in the theory. This property
has been also referred to as "rheonomy" in the literature.
\\
Secondly one writes down such expansion in a form which is
compatible with all the symmetries of the theory, that is:
covariance under $SU(2)$ $R$--symmetry, Lorentz transformations
and reparametrizations of the scalar manifold. Besides one has to
take into account the invariance under the following rigid
rescalings of the fields (and their corresponding curvatures):
\begin{equation}
 (\omega^{ab}, \phi^i , \sigma) \rightarrow
 (\omega^{ab}, \phi^i , \sigma)
 \end{equation}
\begin{equation}
 (V^a, A^\Lambda ) \rightarrow \ell (V^a, A^\Lambda )
 \end{equation}
\begin{equation}
 B_{\mu\nu} \rightarrow \ell^2 B_{\mu\nu}
 \end{equation}
 \begin{equation}
 \psi_A \rightarrow \ell^{1 \over 2}\psi_A
 \end{equation}
 \begin{equation}
 (\lambda^{IA}, \chi_A)
 \rightarrow
 \ell^{-{1 \over 2}}(\lambda^{IA}, \chi_A)
 \label{resc}
 \end{equation}
 Indeed these rescalings and the corresponding ones for the
 curvatures leave invariant the definitions of the curvatures ( in particular
  the F.D.A.) and the Bianchi identities.\\
 Furthermore, the parametrizations of the curvatures
must scale under the $O(1,1)$ duality group discussed in Section 3
in a uniform way.
\par
Taking into account all these constraints, one finds that the
parametrizations of the curvatures of all the fields which solve
the Bianchi identities of the ungauged curvatures (\ref{curve})
have the following form (for notations and definitions see the
text):
\begin{eqnarray}
T^a&=& 0 \nonumber\\
H &=& H_{abc} V^a \wedge V^b \wedge V^c + 4\ii e^{-2\sigma} \pb_A
\g_7 \g_{ab} \chi^A \wedge V^a \wedge V^b \nonumber\\
\widehat{F}^\L &=& \widehat{F}^\L_{ab} V^a \wedge V^b  + 2
e^{\sigma} L^\L_0 \pb_A \g_7 \g_{a} \chi^A \wedge V^a + 2
e^{\sigma} L^\L_{(AB)} \pb^A \g_{a} \chi^B \wedge V^a
+\nonumber\\&&- e^{\sigma} L^\L_I \pb_A \g_{a} \l_I^A \wedge V^a
\nonumber\\
\rho_A &=& \rho_{A|ab} V^a \wedge V^b +
\frac{1}{16}e^{-\sigma}\left(T_{ab}\epsilon_{AB} \g_7
-T_{(AB)|ab}\right)
\left(\g^{abc}-6\delta^{ca}\g^b\right)\p^B\wedge V_c +\nonumber\\
&&+\frac{\ii}{32}e^{2\sigma}H_{abc}\g_7
\left(\g^{dabc}-3\delta^{da}\g^{bc}\right)\p_A\wedge V_d  + \rho_{A (2 \psi)}
+ \rho^{(3f)} \nonumber\\
\nabla\chi_A &=& \nabla_a\chi_A V^a + \frac{\ii}{2} \g^a
\partial_a \sigma \psi_A +
\frac{\ii}{16}e^{-\sigma}\left(T_{ab}\epsilon_{AB} \g_7
+T_{(AB)|ab}\right) \g^{ab}\p^B+\nonumber\\
&&+\frac{1}{32}e^{2\sigma}H_{abc}\g_7 \g^{abc}\p_A  +\nabla\chi^{(3f)} \nonumber\\
\nabla\l^{IA} &=& \nabla_a\l^{IA} V^a + \ii P^{I0}_{i}\partial_a
\phi^i \g_7 \g^a \psi^A -\ii P^{I(AB)}_{i}\partial_a \phi^i \g^a
\psi_B + \frac{\ii}{2}e^{-\sigma}T^I_{ab} \g^{ab}\p^A
  +\nabla \l^{(3f)}\nonumber\\
d\s &=& \partial_a\s V^a +\overline\chi_A\p^A \nonumber\\
P^{I0} &=& P^{I0}_{i}\partial_a\phi^i V^a +\frac 12
\lb^I_A\g_7\p^A
\nonumber\\
P_{I(AB)} &=& P_{I(AB) i}\partial_a\phi^i V^a -\lb_{IA}\p_B
\label{rheo}
\end{eqnarray}
where:
\begin{eqnarray}
\rho_{A(2\psi)}=&&\frac 14 \gamma_{7}\psi_{A}
\overline{\chi}^{C}\gamma^{7}\psi_{C}- \frac 12 \gamma_{a}\psi_{A}
\overline{\chi}^{C}\gamma^{a}\psi_{C}+\frac 12
\gamma_{7}\gamma_{a}\psi_{A}
\overline{\chi}^{C}\gamma^{7}\gamma^{a}\psi_{C}-\frac 18
\gamma_{ab}\psi_{A}
\overline{\chi}^{C}\gamma^{ab}\psi_{C}+\nonumber\\
&&-\frac 18\gamma_{7}\gamma_{ab}\psi_{A}\overline{\chi}^{C}
\gamma^{7}\gamma^{ab}\psi_{C}+\frac 14
\psi_{A}\overline{\chi}^{C}\psi_{C} \label{puzzi}
\end{eqnarray}
and $\rho^{(3f)},\,\nabla\chi^{(3f)},\,\nabla \l^{(3f)}$ denote
terms constructed out in terms of one fermionic vielbein $\p$ and
two spin 1/2 - fermions, namely of the form: $\p \chi\chi,\,\p
\chi \l,\,\p\l\l$ which have not been computed. \footnote{Note
that the flat superspace derivatives along a set of vielbein
$V^a$'s are what in component formalism are called the
supercovariant field--strengths. Indeed, if we project the
curvatures \eq{rheo} along the space--time differentials
$dx^\mu$'s we find, e.g. from the second equation: \be
H_{\mu\nu\rho} = H_{abc} V^a_\mu V^b_\nu V^c_\rho + 4\ii
e^{-2\sigma} \pb_{A[\mu} \g_7 \g_{\nu\rho]} \chi^A \ee and
$\tilde{H}_{\mu\nu\rho} \equiv H_{abc} V^a_\mu V^b_\nu V^c_\rho$
is the supercovariant field--strength. The same observation
applies to all the flat derivatives appearing in equations
\eq{rheo}.}
\par
Let us make a few comments about this solution. An important point
is the presence of the term $\rho_{A(2\psi)}$ in the gravitino
curvature $\rho_A$. Indeed, this term is essential in order to
solve in a consistent way the previous Bianchi identities,
already in the pure supergravity sector, that is setting the
vector multiplets to zero. The pure supergravity Bianchi
identities take, in this case, the following form:
\begin{eqnarray}
\label{torsp}&&R^{ab} V_b-i
\overline{\psi}_A\gamma^a\rho_B\epsilon^{AB}=0 \\
&&\mathcal{D}R^{ab}=0 \\
\label{dHp}&&dH\!+4e^{-2\sigma}d\sigma\,\
\overline{\psi}_A\gamma_7\gamma_a\psi_B\epsilon^{AB}V^a\!+4e^{-2\sigma}
\overline{\psi}_A\gamma_7\gamma_a\rho_B\epsilon^{AB}V^a\!=0
\\ \label{dFp}&&DF+id\sigma e^{\sigma}\,\
\overline{\psi}_A\gamma_7\psi_B\ \epsilon^{AB}-2ie^{\sigma}\,\
\overline{\psi}_A\gamma_7\rho_B\ \epsilon^{AB}=0 \\
\label{dFrp}&&DF^r+id\sigma e^{\sigma}\,\
\overline{\psi}_A\psi_B\,\ \sigma^{rAB}-2ie^{\sigma}\,\
\overline{\psi}_A\rho_B\,\ \sigma^{rAB}=0 \\
\label{drop}&&D\rho_A+\frac{1}{4}R^{ab}\gamma_{ab}\psi_A-\frac{i}{2}\,\
g\,\ \sigma_{rAB}F^r\psi^B=0 \\
\label{dRp}&&DR_A+\frac{1}{4}R^{ab}\gamma_{ab}\chi_A-\frac{i}{2}\,\
g\,\ \sigma_{rAB}F^r\chi^B=0 \\
\label{d2sp}&&d^2\sigma=0
\end{eqnarray}
The necessity of the $\rho_{A(2\psi)}$ term in the gravitino
curvature can indeed be ascertained when one tries to solve the
previous Bianchi identities at the highest level, that is in the
sector containing $\psi \wedge \psi \wedge \psi$ (3 fermionic
vielbeins). Indeed, it is not difficult to verify that the sector
$(V\chi\psi\psi\psi)$ of eq. \eq{dHp} and the sectors
$(\chi\psi\psi\psi)$ of eq.s \eq{torsp}, \eq{dFp} and \eq{dFrp}
do not close unless we add a suitable $\rho_{A(2\psi)}$ as given
in eq. \eq{puzzi} in the gravitino curvature. To arrive at this
result requires a lengthy and cumbersome computation since one
has to use several times the Fierz identities between three-$\psi$
one-forms, discussed in Appendix C.
\par We stress that the
verification of the closure of the Bianchi identities at the
three-$\psi$ level is quite essential; this is to be contrasted
with the fact that, in the analysis of the Bianchi identities, we
neglected the three fermions terms $\rho^{(3f)}$ in the gravitino
curvature containing only one $\psi$, that is terms of the form
$\p\chi\chi$, $\p\l\l$, $\p\l\chi$. Indeed it is well known that
once the Bianchi identities have been satisfied in the highest
sectors with three or two $\psi$, then they automatically close
(on shell) in the sectors containing only one $\p$. In this
sense, since the three-$\p$'s terms in superspace are terms of
the form $\psi\psi\epsilon$ on space--time, and the latter were
neglected in ref. \cite{rom}, we may say that our analysis proves
the consistency of Romans'construction of pure $F(4)$
supergravity also at the three-fermion level.
\par
For the benefit of the reader not familiar with the Superspace
Bianchi identities, we recall that the determination of the
superspace curvatures enables us to write down the space--time
supersymmetry transformation laws. Indeed, from the superspace
point of view, a supersymmetry transformation is a Lie derivative
along the tangent vector:
\begin{equation}
\epsilon = \bar\epsilon^A\,{\vec D}_A
\end{equation}
where the basis tangent vector ${\vec D}_A$ is dual to the
gravitino 1--form:
\begin{equation}
{\vec D}_A \left(\psi^B \right) = \delta^B_A \bf 1
\end{equation}
where $\bf 1$ is the unit in spinor space.
\par
Denoting by $\mu^I$ and $R^I$ the set of 1--forms and 2--form
potential $\Bigl ( V^a,\,\psi_A,\,A^{\Lambda},B \Bigr )$ and of
2--forms and 3--form curvatures $\Bigl (
T^a,\,\rho_A,\,F^{\Lambda},\,H \Bigr )$ respectively, one has:
\begin{equation}
\ell_\epsilon \mu^I =
\left(i_{\epsilon}d\,+\,di_{\epsilon}\right)\mu^I \equiv \left(D
\epsilon\right)^I\,+\,i_{\epsilon}R^I
\end{equation}
where D is the derivative covariant with respect to the $N=2$
Poincar\'e superalgebra and $i_{\epsilon}$ is the contraction
operator along the tangent vector $\epsilon$.
\par
In our case:
\begin{eqnarray}
\left(D \epsilon\right)^a &=& -{\rm
i}\left(\bar\psi_A\gamma^a\epsilon^A \right)\\
\left(D \epsilon_A\right)^{\alpha} &=& \nabla\epsilon^{\alpha}_A\\
\left(D \epsilon\right)^{\Lambda} &=& 0
\end{eqnarray}
(here $\alpha$ is a spinor index)\\
For the 0--forms which we denote shortly as $\nu^I
\equiv\left(q^u,\,\sigma,\,\lambda^{IA},\chi_A \,\right)$ we have
the simpler result:
\begin{equation}
\ell_{\epsilon} = i_{\epsilon}d \nu^I =
i_{\epsilon}\left(\nabla\nu^I\,-\,connection\,terms \right)
\end{equation}
Using the parametrizations given for $R^I$ and $\nabla\nu^I$ and
identifying $\delta_{\epsilon}$ with the restriction of
$\ell_{\epsilon}$ to space--time it is immediate to find the
$N=2$ susy laws for all the fields. The explicit formulae for the
ungauged case are given by the equations (\ref {t1})--(\ref {t6})
of the text.
\par
 Let us now consider the gauged Bianchi identities.
 \\
 In the gauged theory, the curvatures are defined formally as in eqs \eq{curve}
provided we use the gauged connections and vielbeins of the
scalar manifold: $ \hat \Omega = L^{-1} \nabla L $, defined in
equations (\ref {nabla})--(\ref {15}), instead of the ungauged
ones: $ \Omega = L^{-1} d L$. \par Correspondingly, the new gauged
Bianchi identities become:
\begin{eqnarray}
\label{xtors1}&&R^{ab} V_{b}-i
\overline{\psi}_{A}\gamma^{a}\rho_{B}\epsilon^{AB}=0\\
&&\mathcal {D}R^{ab}=0\\
\label{xdHg}&&dH+4e^{-2\sigma}d\sigma\,\
\overline{\psi}_{A}\gamma_{7}\gamma_{a}\psi_{B}\epsilon^{AB}V^{a}+4e^{-2\sigma}
\overline{\psi}_{A}\gamma_{7}\gamma_{a}\rho_{B}\epsilon^{AB}V^{a}=0\\
 \label{xdFg}&&DF^{\Lambda}+id\sigma
e^{\sigma}\overline{\psi}_{A}\gamma_{7}\psi_{B}L^{\Lambda}_{[AB]}+id\sigma
e^{\sigma}\overline{\psi}_{A}\psi_{B}L^{\Lambda}_{(AB)}\nonumber\\
&&-2ie^{\sigma}
\overline{\psi}_{A}\gamma_{7}\rho_{B}L^{\Lambda}_{[AB]}
-2ie^{\sigma}
\overline{\psi}_{A}\rho_{B}L^{\Lambda}_{(AB)}+ie^{\s}L^{\L}_{\ \
I}\pb^A\p^B\hat P^I_{(AB)}+\nonumber\\
&&+ie^{\s}L^{\L}_{\ \ I}\pb^A\g_7\p^B \hat P^I_{[AB]}=0\\
\label{xdrog}&&D\rho_{A}+\frac{1}{4}R^{ab}\gamma_{ab}\psi_{A}-
\frac{i}{2}\sigma^{r}_{AB}(\frac{1}{2}\epsilon^{rst}\hat\mathcal{R}^{st}
+i\gamma_{7}\hat\mathcal{R}_{r0})\psi^{B}=0\\
\label{xdRg}&&D^2\chi_{A}+\frac{1}{4}R^{ab}\gamma_{ab}\chi_{A}-
\frac{i}{2}\sigma^{r}_{AB}(\frac{1}{2}\epsilon^{rst}\hat\mathcal{R}^{st}
+i\gamma_{7}\hat\mathcal{R}_{r0})\chi^{B}=0\\
&&d^{2}\sigma=0\\
\label{xd2lg}&&D^{2}\lambda^{I}_{A}+\frac{1}{4}R^{ab}\gamma_{ab}\lambda^{I}_{A}
-\frac{i}{2}\sigma^{r}_{AB}(\frac{1}{2}\epsilon^{rst}
\hat\mathcal{R}^{st}+i\gamma_{7}\hat\mathcal{R}_{r0})\lambda^{IB}
-\hat\mathcal{R}^{I}_{J}\lambda^{J}_{A}=0\\
\label{xdPrg}&&D\hat P^{I}_{r}= \left(L^{-1} \tilde {\cal F} L
\right)^I_r\\
\label{xdP0g}&&D\hat P^{I}_{0}= \left(L^{-1} \tilde {\cal F} L
\right)^I_0
\end{eqnarray}
where all the "hatted" quantities have been defined in the text
and the covariant derivatives are now covariant also with respect
to the gauge group $SU(2)\otimes \mathcal {G}$. As it happens in
all gauged supergravities, the new solution of the Bianchi
Identities differs from the old one in the following aspects:
\begin{enumerate}
\item{The vector field--strengths are now non abelian.}
\item{
The derivatives have to be made covariant also with respect to the
gauge group, so that the non abelian field--strengths appearing
on the l.h.s. of the old parametrizations \eq{rheo} now contain
the "hatted" connections and vielbeins of the scalar manifold.}
\item{The parametrizations of the fermionic curvatures
$ \rho_A,$, $\nabla \chi_A$, $\nabla \lambda^{IA}$ contain extra
terms $S_{AB}\gamma_a \psi V^a,\,N_{AB}\psi^B,\,M^I_{AB}\psi^B$
which are proportional to the gauge coupling constants $g,g'$.}
\end{enumerate}

As it has been explained in the text, there is however a more
general solution of the B.I. involving a second "gauging " in
terms of the mass parameter $m$ so that the previous fermionic
shifts can also acquire extra terms proportional to $m$. The
computation of the complete shifts is actually quite cumbersome
since we have to use several times the relevant Fierz--identities
quoted in Appendix C and further decompose the relevant structures
in $SU(2)$ irreducible fragments. The final result for the shifts
is given in the text (see equations (\ref{S})--(\ref{fine}). The
resulting supersymmetry transformation laws on space--time can be
finally obtained as explained in the ungauged case and are given
by eq. \eq{newsusy} of the text.

\section*{Appendix B: The Lagrangian from the geometric approach}
\setcounter{equation}{0} \label{appendiceB}
\addtocounter{section}{1}
 In Appendix A we have
outlined how to recover the supersymmetry transformation laws for
the physical fields of the matter coupled $F(4)$ theory from the
solution of Bianchi identities in superspace. Since the closure of
Bianchi identities is true only on the mass-shell, the equations
of motion for all the fields are also implicitely given, and from
them one could in principle reconstruct the Lagrangian. However,
this procedure would be quite cumbersome. We therefore prefered
to work out the space--time Lagrangian from a geometric
Lagrangian in superspace, whose construction in the geometric
(rheonomic) approach is straightforward. This appendix gives a
short account of its derivation.
\par
In the geometric approach the superspace action for the theory is
a six-form in superspace \footnote{The superspace we are
considering, ${\cal M}^{6|16}$, contains 6 space--time directions
and 16 ($N=2$) fermionic directions.} integrated on a six
dimensional (bosonic) hypersurface ${\cal M}^6$ locally embedded
in ${\cal M}^{6|16}$.: \be {\cal A} = \int_{{\cal M}^6\subset
{\cal M}^{6|16}} {\cal L} \ee
 It contains the fields of the theory
through external forms on ${\cal M}^{6|16}$, using only
dif\-feo\-mor\-phism-invariant operations of external algebra,
namely the exterior derivative $d$ and the wedge product $\wedge$
(we therefore never introduce the Hodge duality operator, which
depends on the choice of the hypersurface of integration). We then
make use of a generalized variational principle ($\delta {\cal A}
= 0$), which provides superspace equations of motions that are
4-form, 5-form or 6-form equations independent from the particular
hypersurface ${\cal M}^6\subset {\cal M}^{6|16}$ on which we
integrate. These superspace equations of motion can be analyzed
along the $p$--form basis. The components of the equations along
bosonic vielbeins give the differential equations for the fields
which, identifying ${\cal M}^{6}$ with space--time, are the
ordinary equations of motion of the theory. The components of the
same equations along $p$-forms containing at least one gravitino
("outer components") instead, according to the priciple of
rheonomy, must be all expressed in terms of the supercovariant
internal components (components along the bosonic vielbeins
basis). Actually if we have already solved the Bianchi identities
this requirement is equivalent to identify the outer components of
the curvatures obtained from the variational principle with those
obtained from the Bianchi identities.
\par
There are simple rules which can be used in order to write down
the most general Lagrangian compatible with this requirement.\\
 Actually one writes down the most general
6-form as a sum of terms with indeterminate coefficients in such a
way that $\cal L$ be a scalar with respect to all the symmetry
transformations of the theory. In order to avoid the use of the
Hodge operator (which would destroy the independence of the
variational equations from the particular hypersurface of
integration) the kinetic terms of the Lagrangian have been written
in first--order formalism. Specifically one introduces auxiliary
0--forms namely ${\bf{H}}_{abc}$, ${\bf{F}}^{\Lambda}_{ab}$,
${\bf P}^{I AB}_a$, ${\bf \Sigma_a }$, whose variational equations
identify them with ${{H}}_{abc}$, $\hat{F}^{{\Lambda}}_{ab}$, $
P^{I AB}_i \partial_a \phi^i \equiv P^{I AB}_a$, $\partial_a
\sigma $ appearing in equations \eq{rheo} of Appendix A. Also the
spin connection $\omega^{ab}$ has to be treated as an independent
field: indeed the terms in the Lagrangian containing explicitely
the torsion $T^a$ have been chosen in such a way that the
equation of motion of $\omega^{ab}$ gives $T^a = 0$.
\par
Varying the action and comparing the outer equations of motion
with the actual solution of the Bianchi identities one then fixes
all the undetermined coefficients except the coefficients of terms
that are proportional to $V^a\wedge V^b \wedge \cdots V^f
\epsilon_{abcdef}$. Indeed, after variation, these last terms do
not contain any fermionic vielbein $\psi$ and appear therefore in
the space--time equations of motion.These undetermined
coefficients, however, can be retrieved by the request that the
superspace lagrangian be invariant under  supersymmetry
transformation, that is by implementing the condition: \be
\delta_\epsilon{\cal L}=i_\epsilon d{\cal L}=0 \ee where
$i_\epsilon$ denotes contraction of the generator of
supersymmetry $\epsilon=\epsilon^A Q_A$ on the form.
\par Let us perform the steps previously indicated. The most
general 6-form Lagrangian, up to four--fermions terms, has the
following form: \ba
&&\mathcal{L}^{(g,g',m)}_{(D=6,N=2)}=\nonumber\\
%%%%%%%%%%%%%%%%%%%%%%%%%%%%%%%%%%%%%%%%%%%%%%%%%%%%%
%%%%%%%%%lagrangiana cinetica %%%%%%%%%%%%%%%%%%%%%
%%%%%%%%%%%%%%%%%%%%%%%%%%%%%%%%%%%%%%%%%%%%%%%%%%%%
&&=\mathcal{R}^{ab}\wedge V^c\cdots\wedge V^f\epsilon_{a\cdots f}
+\left(p_1
\overline{\lambda}^I_A\gamma^a\nabla\lambda^A_I+p_2\overline{\chi}_A\gamma_a
\nabla\chi^A\right)\wedge V^b\cdots\wedge V^f\epsilon_{a\cdots f}+
\nonumber\\
&&+ a_1\overline{\psi}_{A}\g_7\gamma^{abc}\rho^{A}\wedge V^a
\cdots\wedge V^c+\Bigl[b_3{\bf \Sigma_a}\left(d\sigma
-\bar\chi_{A}\p^A\right)+\nonumber\\
&&+  b_1{\bf {P}}^{I0}_{a}\left(\widehat{P}_{I0}-\half
\bar\lambda_{IA}\g_7\p^A\right)+ b_2{\bf
{P}}^{I(AB)}_{a}\left(\widehat{P}_{I(AB)}
+\bar\lambda_{IA}\p_B\right)\Bigr]\wedge V_b \cdots \wedge V_f
\epsilon^{a\cdots f}+\nonumber\\
&&-\frac{1}{12}\left[b_1{\bf  P}^I_{m}{\bf P}_I^{m}+ b_2{\bf
P}^{I(AB)}_{m}{\bf P}_{I(AB)}^{m}+ b_3 {\bf
\Sigma_m}{\bf\Sigma^m}\right]V_a\cdots\wedge V_f
\epsilon^{a\cdots f}+\nonumber\\
&&+d_1e^{-2\sigma}\mathcal{N}_{\Lambda\Sigma}\Bigl[
{\bf{F}}^{\Lambda}_{ab}\Bigl(\widehat{F}^\Sigma-2 e^\s
L^\Sigma_0\pb_A\g_7\g_\ell\chi^A\wedge V^\ell-2e^\s
L^\Sigma_{(AB)}\pb^A\g_\ell\chi^B\wedge V^\ell +\nonumber\\
&&+ e^\s L^\Sigma_I \pb_A \g_\ell \l^{IA}\wedge V^\ell\Bigr)\wedge
V_c \cdots \wedge V_f \epsilon^{a\cdots f}- \frac{1}{30} {\bf
F}^{\Lambda}_{\ell m}{\bf F}^{\Sigma\ell m}V^a \cdots \wedge V^f
\epsilon_{a\cdots f}\Bigr]+\nonumber\\
&&+ f_1 e^{4\s}{\bf H}_{abc}\Bigl[H-4\ii e^{-2\s}\pb_A\g_7
\g_{\ell m }\chi^{A}\wedge V^\ell \wedge V^m \Bigr]\wedge V_d
\cdots \wedge
V_f\epsilon^{a\cdots f}+\nonumber\\
&&- \frac{1}{40}f_1 e^{4\s}{\bf H}_{\ell m n}{\bf H}^{\ell m n}V_a
\cdots \wedge V_f \epsilon^{a\cdots f}+
\nonumber\\
%%%%%%%%%%%%%%%%%%%%%%%%%%%%%%%%%%%%%%%%%%%%%%%%%%%%%
%%%%%%%%%lagrangiana di pauli %%%%%%%%%%%%%%%%%%%%%
%%%%%%%%%%%%%%%%%%%%%%%%%%%%%%%%%%%%%%%%%%%%%%%%%%%%
&&+ \left[g_1 d\s \pb_A \g_{ab} \chi^A + g_2 \widehat{P}^I \pb_A
\g_7 \g_{ab} \l^A_I + g_3 \widehat{P}^I_{(AB)} \pb^A \g_7 \g_{ab}
\l^B_I\right]\wedge V_c \cdots \wedge V_f \epsilon^{a\cdots f}
+\nonumber\\
&&+e^{-\s}\mathcal{N}_{\L\Sigma}\hat{F}^{\L} \Bigl[ \left(h_1
L^{\Sigma}_0\pb_A \g_{ab}\p^{A}+
 h_2 L^{\Sigma}_{(AB)}\pb^{A}\g_7 \g_{ab}\p^{B}\right) \wedge
 V^a\wedge V^b + \nonumber\\
&&+\left(h_3 L^{\Sigma}_0\pb_A \g_7\g^{abc}\chi^{A}+ h'_3
L^{\Sigma}_{(AB)}\pb^A \g^{abc}\chi^B+ h_4 L^{\Sigma}_I\pb_A
\g^{abc}\l^{IA}\right)\wedge V^d \cdots \wedge V^f
\epsilon_{a\cdots f}
+\nonumber\\
&&+\Bigl( h_5 L^{\Sigma}_0\lb_{IA} \g_7\g^{ab}\l^{IA}+ h'_5
L^{\Sigma}_{(AB)}\lb_{I}^A\g^{ab}\l^{IB}+ h_6
L^{\Sigma}_0\overline{\chi}_{A} \g_7\g^{ab}\chi^{A}+
\nonumber\\
&&+ h'_6 L^{\Sigma}_{(AB)}\overline{\chi}^A\g^{ab}\chi^{B}\Bigr)
\wedge V^c \cdots \wedge V^f \epsilon_{a\cdots f}\Bigr]
+\nonumber\\
 &&+ e^{2\s}H\wedge
\Bigl[p'_1 \pb_A \g_{a}\p^{A}\wedge V^a+
 p'_2 \pb_{A} \g_{ab}\chi^{A}\wedge V^a\wedge V^b+\nonumber\\
  &&+ p_5 \overline{\chi}_{A} \g_{abc}\chi^{A} V^a\cdots \wedge V^c\Bigr]+\nonumber\\
&& +T_a \wedge \Bigl[\left(\ell_1 \lb_{I}^A\g_7\g_{bcd}\l^{IA}+
\ell_2 \overline{\chi}_{A} \g_7\g_{bcd}\chi^{A}\right)V^a
\cdots\wedge V^d + \nonumber\\
&&+ \ell_3 \pb_A \g_7 \g_b \p^A \wedge V^a \wedge
V^b\Bigr] +\nonumber\\
%%%%%%%%%%%%%%%%%%%%%%%%%%%%%%%%%%%%%%%%%%%%%%%%%%%%%
%%%%%%%%%lagrangiana di chern-simons %%%%%%%%%%%%%%%
%%%%%%%%%%%%%%%%%%%%%%%%%%%%%%%%%%%%%%%%%%%%%%%%%%%%
&& + k B \wedge \Bigl[\eta_{\L\Sigma}\left(\widehat{F}^{\L} + \ii
e^\s L^\L_0 \pb_A \g_7 \p^A + \ii e^\s L^\L_{(AB)} \pb^A \p^B
\right)\wedge \nonumber\\
&&\wedge  \left(\widehat{F}^{\Sigma} + \ii e^\s L^\Sigma_0 \pb_C
\g_7 \p^C + \ii e^\s L^\Sigma_{(CD)} \pb^C \p^D
\right)+\nonumber\\
&&+ mB \delta_\L^0 \wedge \left(\widehat{F}^{\L} + \ii e^\s L^\L_0
\pb_A \g_7 \p^A + \ii e^\s L^\L_{(AB)} \pb^A \p^B \right)
+\frac{1}{3}m^2B\wedge B\Bigr] + \nonumber\\
%%%%%%%%%%%%%%%%%%%%%%%%%%%%%%%%%%%%%%%%%%%%%%%%%%%%%
%%%%%%%%%lagrangiana di gauging %%%%%%%%%%%%%%%%%%%
%%%%%%%%%%%%%%%%%%%%%%%%%%%%%%%%%%%%%%%%%%%%%%%%%%%%
&& +\Bigl[\delta_1 \pb_A \g^{ab} \overline{S}^{AB} \p_B  + \left(
\delta_2 \pb_A \g^{a} \overline{N}^{AB} \chi_B + \delta_3 \pb_A
\g^{a} \overline{M}_I^{AB} \l^I_B\right) \wedge V^b
\Bigr]V^c\cdots \wedge V^f \epsilon_{a\cdots f}+ \nonumber\\
&&+ \delta_4\left( \overline{\chi}^A X_{AB} \chi^B +
\overline{\chi}^A Y^I_{AB} \lambda_I^B +  \lb^I_A
Z^{AB}_{IJ}\lambda^J_B \right)V^a \cdots \wedge V^f
\epsilon_{a\cdots f}+\nonumber\\
&& -\mathcal{W}(\sigma , \, \phi^i;g,g',m)V^a \cdots \wedge V^f
\epsilon_{a\cdots f}+\mathcal{L}_{\mbox{\tiny{4 fermions
}}}.\label{lagrheo}\ea The mass matrices for the spin 1/2
fermions, $X_{AB}$, $Y^I_{AB}$, $Z^{AB}_{IJ}$, were computed
through the request of supersymmetry invariance of the superspace
lagrangian and are given by \eq{massespinori}.
\par
The coefficients appearing in \eq{lagrheo} have been found, from
the superspace variational equations, to be:
 \ba a_1 &=& -8 \, ; \ b_1 = -\frac 25 \, ; \
b_2 = \frac 15 \, ; \ b_3 = \frac 85 \,
; \ d_1 = -\frac 12 \, ; \ f_1 = \frac 14 \nonumber\\
p_1 &=& -\frac{\ii}{10} \, ; \ p_2 = -\frac 85 \ii \, ; \ k =
\frac 32 \, ; \ h_1 = 6 \,
; \ h_2 = -6 \, ; \ h_3 = -\frac 23  \nonumber\\
 h'_3 &=& -\frac 23 \, ; \ h_4 = -\frac 13 \, ; \ h_5 = \frac{\ii}{16} \, ;
\ h'_5 = \frac{\ii}{16} \,
; \ h_6 = \frac{\ii}{2} \, ; \ h'_6 = -\frac{\ii}{2} \nonumber\\
 p'_1 &=& 3\ii \, ; \ p'_2 = -6 \, ; \ p_5 = -2\ii \, ;
\ g_1 = -4 \,
; \ g_2 = -\frac 12 \, ; \ g_3 = -\frac 12 \nonumber\\
\ell_1 &=& 3 \, ; \ \ell_2 = -48 \, ; \ \ell_3 = -12 \, ; \
\delta_1 = 4\ii \,
; \ \delta_2 = \frac{16}{5}\ii \, ; \ \delta_3 = -\frac{\ii}{5} \nonumber\\
\delta_4 &=& \frac{2}{15}. \ea
\par
In order to obtain the space--time Lagrangian the last step to
perform is the restriction of the 6--form Lagrangian from
superspace to space--time. This corresponds to restrict all the
terms to the particular hypersurface ${\cal M}^6$ with $\theta =
0\,,\,d \theta = 0$ . In practice one first goes to the second
order formalism by identifying the auxiliary 0--form fields as
explained before. Then one expands all the forms along the
$dx^{\mu}$ differentials and restricts the superfields to their
lowest ($\theta = 0$) component. Finally the coefficients of:
\begin{equation}
dx^{\mu_1}\wedge \cdots \wedge
dx^{\mu_6}\,=\,{\epsilon^{\mu_1\cdots\mu_6}\over \sqrt g}\left(
\sqrt g d^6x \right)
\end{equation}
give the Lagrangian density written in section $3$. The overall
normalisation of the space--time action has been chosen such as to
be the standard one for the Einstein term.
%%%%%%%%%%%%%%%%%%%%%%%%%%%%%%%%%%%%%%%%%%%%%%%
%%%%%%%%%%%%%%%%%%%%%%%%%%%%%%%%%%%%%%%%%%%%%%%
%%%%%%%%%%%%%%%%%%%%%%%%%%%%%%%%%%%%%%%%%%%%%%%
\section*{Appendix C: $D=6$, $N=2$ Fierz identities for pseudo-Majorana spinors}
\setcounter{equation}{0} \label{appendiceC}
\addtocounter{section}{1}
All the Fierz identities can be derived
from the following fundamental bilinear identity:
{\setlength\arraycolsep{1pt}\begin{eqnarray}\psi_B\overline{\psi}_A=&&\frac{1}{8}\overline{\psi}_A\psi_B
1-\frac{1}{8}\overline{\psi}_A\gamma^{7}\psi_B
\gamma_7+\frac{1}{8}\overline{\psi}_A\gamma^{a}\psi_B
\gamma_a+\frac{1}{8}\overline{\psi}_A\gamma^{7}\gamma^a\psi_B
\gamma_7\gamma_a+\nonumber\\
\label{paolo}&&-\frac{1}{16}\overline{\psi}_A\gamma^{ab}\psi_B
\gamma_{ab}+\frac{1}{16}\overline{\psi}_A\gamma^{7}\gamma^{ab}\psi_B
\gamma_7\gamma_{ab}-\frac{1}{48}\overline{\psi}_A\gamma^{abc}\psi_B
\gamma_{abc}\end{eqnarray}} Using (\ref{paolo}) and the symmetry
properties of the gamma--matrices it is immediate to check the
following quadri-linear identity:
\begin{eqnarray}\label{claudia}
\overline{\psi}_{A}\gamma_{a}\psi_{B}\overline{\psi}_{C}\gamma^{7}\gamma^{a}\psi_{D}\epsilon^{AB}\epsilon^{CD}=0.
\end{eqnarray}
In order to obtain three-linear identities, we write down all the
possible three-linear terms with no explicit Lorentz index and
with a given $SU(2)$ tensor structure. We have:
\begin{eqnarray*}
&&A^1_A=\gamma_{7}\psi_{A}\overline{\psi}_{B}\gamma^{7}\psi_{C}\epsilon^{BC}\\
&&A^2_A=\gamma_{a}\psi_{A}\overline{\psi}_{B}\gamma^{a}\psi_{C}\epsilon^{BC}\\
&&A^3_A=\gamma_{7}\gamma_a\psi_{A}\overline{\psi}_{B}\gamma^{7}\gamma^a\psi_{C}\epsilon^{BC}\\
&&A^4_A=\gamma_{ab}\psi_{A}\overline{\psi}_{B}\gamma^{ab}\psi_{C}\epsilon^{BC}\\
&&S^1_A=\psi^B\overline{\psi}_A\psi_B\\
&&S^2_A=\gamma_{abc}\psi^B\overline{\psi}_A\gamma^{abc}\psi_B\\
&&S^3_A=\gamma_{7}\gamma_{ab}\psi^B\overline{\psi}_A\gamma^{7}\gamma^{ab}\psi_B\\
&&S^{1}_{(ABC)}=\psi_{(A}\overline{\psi}_B\psi_{C)}\\
&&S^{2}_{(ABC)}=\gamma_{abc}\psi_{(A}\overline{\psi}_B\gamma^{abc}\psi_{C)}\\
&&S^{3}_{(ABC)}=\gamma_{7}\gamma_{ab}\psi_{(A}\overline{\psi}_B\gamma^{7}\gamma^{ab}\psi_{C)}.
\end{eqnarray*}
A group-theoretical analysis. together with numerical Mathematica
computations, gives the following Fierz identities:
\begin{eqnarray}
\label{mbuto1}&&A^3_A=A^2_A\\
\label{mbuto2}&&A^4_A=6A^1_A-4A^2_A\\
\label{mbuto3}&&S^1_A=\frac{1}{2}A^2_A-\frac{1}{2}A^1_A\\
\label{mbuto4}&&S^3_A=-3A^1_A-3A^2_A\\
\label{mbuto5} &&S^{3}_{(ABC)}=6S^{1}_{(ABC)}\\
\label{mbuto6}&&S^{2}_{(ABC)}=-24S^{1}_{(ABC)}.
\end{eqnarray}
In particular, for the purpose of solving the vectors Bianchi
identities at the three $\p$-level , one can derive from
equations (\ref{mbuto1})--(\ref{mbuto4}) the following useful
identity:
\begin{equation}
\label{pucci}4\overline{\chi}_{(A}\gamma_{a}\psi_{B)}\overline{\psi}_{C}\gamma^{A}\psi^{C}
-6\overline{\chi}_{C}\psi^{C}\overline{\psi}_{(A}\psi_{B)}
+\overline{\chi}_{C}\gamma_{7}\gamma_{ab}\psi^{C}\overline{\psi}_{(A}\gamma^{7}\gamma^{ab}\psi_{B)}=0.
\end{equation}
In the sector of the three-linear objects constructed out three
$\p$ and with one free Lorentz index, we just quote the
structures which are $SU(2)$ vectors entering in the computation
of the Bianchi identities. They are:
\begin{eqnarray*}
&&B^{1a}_A=\gamma_{7}\psi_{A}\overline{\psi}_{B}\gamma^{7}\gamma^a\psi_{C}\epsilon^{BC}\\
&&B^{2a}_A=\gamma^{7}\gamma^a\psi_{A}\overline{\psi}_{B}\gamma_{7}\psi_{C}\epsilon^{BC}\\
&&B^{3a}_A=\gamma_{b}\psi_{A}\overline{\psi}_{B}\gamma^{ab}\psi_{C}\epsilon^{BC}\\
&&B^{4a}_A=\gamma^{ab}\psi_{A}\overline{\psi}_{B}\gamma_{b}\psi_{C}\epsilon^{BC}\\
&&B^{5a}_A=\gamma^{7}\gamma^{ab}\psi_{A}\overline{\psi}_{B}\gamma_{7}\gamma_b\psi_{C}\epsilon^{BC}\\
&&B^{6a}_A=\gamma^{abc}\psi_{A}\overline{\psi}_{B}\gamma_{bc}\psi_{C}\epsilon^{BC}\\
&&B^{7a}_A=\psi_{A}\overline{\psi}_{B}\gamma^a\psi_{C}\epsilon^{BC}\\
\end{eqnarray*}
Again, a group theoretical and numerical analysis give the
following three relations among the seven quantities:
\begin{eqnarray}
&&B^{3a}_A=-B^{1a}_A+\frac{1}{2}B^{5a}_A+\frac{1}{4}B^{6a}_A+\frac{3}{2}B^{7a}_A\nonumber\\
&&B^{4a}_A=-B^{1a}_A-B^{5a}_A-B^{7a}_A\nonumber\\
\label{patata}&&B^{2a}_A=B^{1a}_A+\frac{1}{2}B^{5a}_A-\frac{1}{4}B^{6a}_A-\frac{1}{2}B^{7a}_A
\end{eqnarray}
The quoted relations \eq{mbuto1}--\eq{mbuto6} and \eq{patata} are
all we need for the solution of Bianchi identities and the
construction of the Lagrangian.
%%%%%%%%%%%%%%%%%%%%%%%%%%%%%%%%%%%%%%%%%%%%%%%%%%%%%%%%%%%%%%%
%%%%%%%%%%%%%%%%%%%%%%%%%%%%%%%%%%%%%%%%%%%%%%%%%%%%%%%%%%%%%%%

\end{document}